\documentclass[journal]{IEEEtran}
\usepackage{amsmath,amssymb,amsfonts}
\usepackage{algorithm}
\usepackage{algorithmic}
\usepackage{graphicx}
\usepackage{textcomp}
\usepackage{subfig}

\def\BibTeX{{\rm B\kern-.05em{\sc i\kern-.025em b}\kern-.08em
    T\kern-.1667em\lower.7ex\hbox{E}\kern-.125emX}}

\usepackage{fancyhdr}
\fancypagestyle{plain}{
  \fancyhf{}
  
  \fancyfoot[C]{\footnotesize This work has been submitted IEEE Transactions on Quantum Engineering for possible publication. Copyright may be transferred later.}
}
\begin{document}

\title{Heuristics for Shuttling Sequence Optimization for a Linear Segmented Trapped-Ion Quantum Computer}

\author{
J.~Durandau,
C.~A.~Brunet,
F.~Schmidt-Kaler,
U.~Poschinger,
F.~Mailhot,
and Y.~Bérubé-Lauzière%
\thanks{J. Durandau, C. A. Brunet, F. Mailhot and Y. Bérubé-Lauzière are with the Institut quantique and the Département de génie électrique et de génie informatique, Université de Sherbrooke, Sherbrooke, Québec, J1K 2R1, Canada.}%
\thanks{F. Schmidt-Kaler and U. Poschinger are with QUANTUM, Institute of Physics, Johannes Gutenberg University, 55128 Mainz, Germany.}
}


\maketitle

\begin{abstract}

An algorithm for the generation of shuttling sequences is necessary for the operation of a linear segmented ion-trap quantum computer. The present work provides an implementation of an algorithm that produces sequences proved to be optimal for circuits with a quantum Fourier transform-like structure. Such optimality was proved in previous work of our group.
We first present an approach for qubit mapping, i.e. determining the initial ordering 
of the ions, termed the \emph{common ion order}, and develop a heuristic algorithm
for its implementation. We explain how this heuristic is integrated in the shuttling sequence generation algorithm described in the previous work. The results show the increased performance of the heuristic in terms of reducing the number of required shuttling operations.
The number of ion displacements required exhibits a polynomial increase in terms of the number of qubits, such that these operations become the main contribution to the overall resource cost.
Furthermore, we show that multiple zones for gate interactions can reduce the amount of qubit register reordering.
\end{abstract}

\section*{Introduction}
\label{Intro}
\IEEEPARstart{A}{s} the quest for more powerful quantum computing platforms continues, the need for increasing the number of qubits grows accordingly. This is a challenge for all types of quantum computers, in particular for linear ion-trap quantum computers, in which qubits are encoded in the electronic states of ions, as the operational fidelities generally decrease with the size of the qubit register~\cite{ionDephasingGate}. To overcome this limitation, the QCCD architecture has been proposed~\cite{Kielpinski2002ArchitectureFA}, which is based on segmented ion traps that are dynamically reconfigurable arrays of multiple traps, each containing groups of a few electrically confined atomic ions, which are referred to as \textit{crystals}~\cite{Figgatt2019,Schindler2013}. This facilitates the implementation of high-fidelity quantum gates as the number of ions locally confined in a trap segment remains small. In such an architecture, gates are  performed using laser beams (microwaves can also be used) directed at a specific trap segment, to be called here a laser interaction zone (LIZ). More precisely, within the LIZ, either one or two ions are held for single- and two-qubit operations.

As gate operations can only be performed in a LIZ, ions must be shuttled there to perform quantum gates. Hence, the placement of the ions in the trap must be dynamically reorganized to execute quantum circuits. Such spatial reorganization is carried out through \textit{shuttling operations}~\cite{Pino2021,Hilder2022}. 

The reorganization of ions incurs costs in the form of timing overhead and possibly shuttling-induced increased error rates. A trapped ion platform therefore ultimately requires a \emph{shuttling algorithm} to produce the most optimized reorganization sequence possible. This is an instance of the qubit mapping problem, proven to be NP-complete~\cite{QubitAllocation,NP_Complet_Problem}. 

At the start of a shuttling algorithm, an \emph{initial qubit ordering}, \textit{i.e.} mapping from algorithmic qubits to stored ions, has to be established.
Concretely, this means assigning a fixed qubit index to each ion stored at a given initial location in the trap. The initial ordering is of great importance because subsequent rearrangements of the ions via shuttling operations entail multiple ion permutations in the LIZ, leading to additional shuttling costs. The computational complexity of the problem of finding an optimal initial ordering grows exponentially with the number of qubits and the size of the architecture, as for $N$ qubits there are on the order of $N!$ possible different assignments of qubits to ions~\cite{1article}. Therefore, a suitable heuristic is needed to find an acceptable initial ordering. Such a heuristic, which exploits circuit structure and extends previous work~\cite{1article}, is presented here.

A linear segmented trap architecture with a single LIZ is first considered here, such as the segmented ion trap quantum computer developed in Mainz~\cite{shuttlingIonTrapQuantum}. This is then extended to a multi-LIZ architecture (see ~\cite{helios} for an example of such an architecture). To implement gates in a circuit, ions need to be exchanged between crystals through operations that split and merge crystals. Such operations are costly in time and error~\cite{Hilder2022,Ruster2014}. Another constraint may be that split and merge operations can be calibrated and executed best in the LIZ, which increases the complexity of the shuttling sequences.
A further constraint may be that only single ions can be exchanged between crystals but not entire crystals. An algorithm for generating adequate shuttling sequences on a linear segmented ion-trap platform must take all such constraints into account.
The initial qubit mapping being critical for minimizing the shuttling sequence cost, we hereby present a heuristic for an initial ordering integrated into a shuttling algorithm, such that shuttling sequences can be generated within an acceptable calculation time and with minimal amounts of shuttling operations, while satisfying the constraints imposed by the architecture.

This paper is organized as follows: Being a foundational prerequisite to the present work, Sect.~\ref{COSTSECTION} reviews the cost metric that quantifies the operational overhead of a shuttling sequence, as introduced in our previous work~\cite{1article}. Sect.~\ref{Model} discusses the relevant details of the linear segmented shuttling-based ion-trap quantum computer. Sect.~\ref{sec:CIO} details the heuristic used to determine acceptable initial
qubit mappings 
and the implementation of the shuttling sequence generating algorithm. The methodology and circuits used for testing the algorithm are described in Sect.~\ref{sec:Imp}. Results obtained with the heuristic are presented and discussed in Sect.~\ref{sec:ResultsCIO}. Then, Sect.~\ref{sec:IonReorg} introduces a different type of heuristic to solve the problems identified in Sect.~\ref{sec:ResultsCIO}, while Sect.~\ref{sec:Displ} discusses the ultimate limit induced by the cost of the displacement of ion crystals. Finally. Sect.~\ref{sec:ML} describes gains obtained by considering a multi-LIZ architecture as regards crystal displacements.

\section{Cost model}
\label{COSTSECTION}

To quantify how well a shuttling sequence generated by an algorithm is adapted to a given quantum circuit, a metric is needed that ideally does not depend on the number of gates or structure of the circuit. A metric that closely fulfills these requirements is the \textit{circuit fit}, to be denoted by $C$, that we have introduced previously~\cite{1article}, and which is defined as the mean cost per gate as
\begin{equation}
C \equiv \mathrm{Circuit~fit} = \frac{\mathrm{Total~cost}}{\mathrm{Number~of~gates}} .
\label{eq:CircFit}
\end{equation}
The total cost used in the circuit fit is the sum of the costs for each shuttling operation in a shuttling sequence, such as moving an ion between segments (displacement cost). However, other sums or cost definitions may be used to focus on specific aspects or some constraints as will be seen later, since some operations can be optimized by the algorithm with an appropriate cost definition. 

The total cost appearing in the circuit fit that will be considered to start with is the number of crystals split and merge operations in a shuttling sequence, which are the costliest operations as per Table~\ref{tabcost}. Such a total cost accounts at a coarser level of detail for the interactions between the algorithm and the circuits while leaving out the topological constraints. As ion exchanges represent the unique way to move ions in different crystals for two-qubit gates, the reduction of the number of splits and merges leads to a reduction of other types of operations, as the less splits and merges occur, the less the crystals have to move. Later in Sect.~\ref{sec:Displ}, the total cost will account for the costs of all operations appearing in Table~\ref{tabcost}. This will allow a deeper analysis of the limitations of the linear architecture.

\begin{table}[]
    \centering
    \caption{Example of the cost taking into account the increase of the mean phonon number associated with each type of operation in the Mainz quantum computer~\cite{shuttlingIonTrapQuantum}.
    A displacement corresponds to moving a crystal from one segment to one of its immediately neighboring segments; a rotation means rotating a crystal within a segment, a split means breaking up within the LIZ a crystal into its two constituent ions, and a merge means joining within the LIZ two ions to form a crystal.}
    \begin{tabular}{c|c}
         Command & Cost (Mean \# phonons) \\
         \hline
         Displacement & 2 \\
         Rotation & 10 \\
         Split/Merge & 50 \\
         \hline
    \end{tabular}
    \label{tabcost}
\end{table}

The independence of the circuit fit on the number of gates lies in that it is the mean cost of executing one gate of the circuit on a specific quantum computer architecture. The circuit fit allows measuring how the cost of implementing a circuit changes on a given architecture as the number of ions increases. It also allows comparing the cost of implementation of two different circuits. 
Most interesting is the dependence of the circuit fit on the number of ions. Convergence or divergence of the circuit fit indicates how the structure of the circuit, the quantum computer's architecture topology, and the shuttling algorithm interact. A divergent circuit fit highlights that the disorganization of the ions grows as the number of ions increases, with disorganization 
meaning
here
the distance between two ions that must be brought in interaction.

\begin{figure}[h!]
\centering
\includegraphics[width=0.45\textwidth]{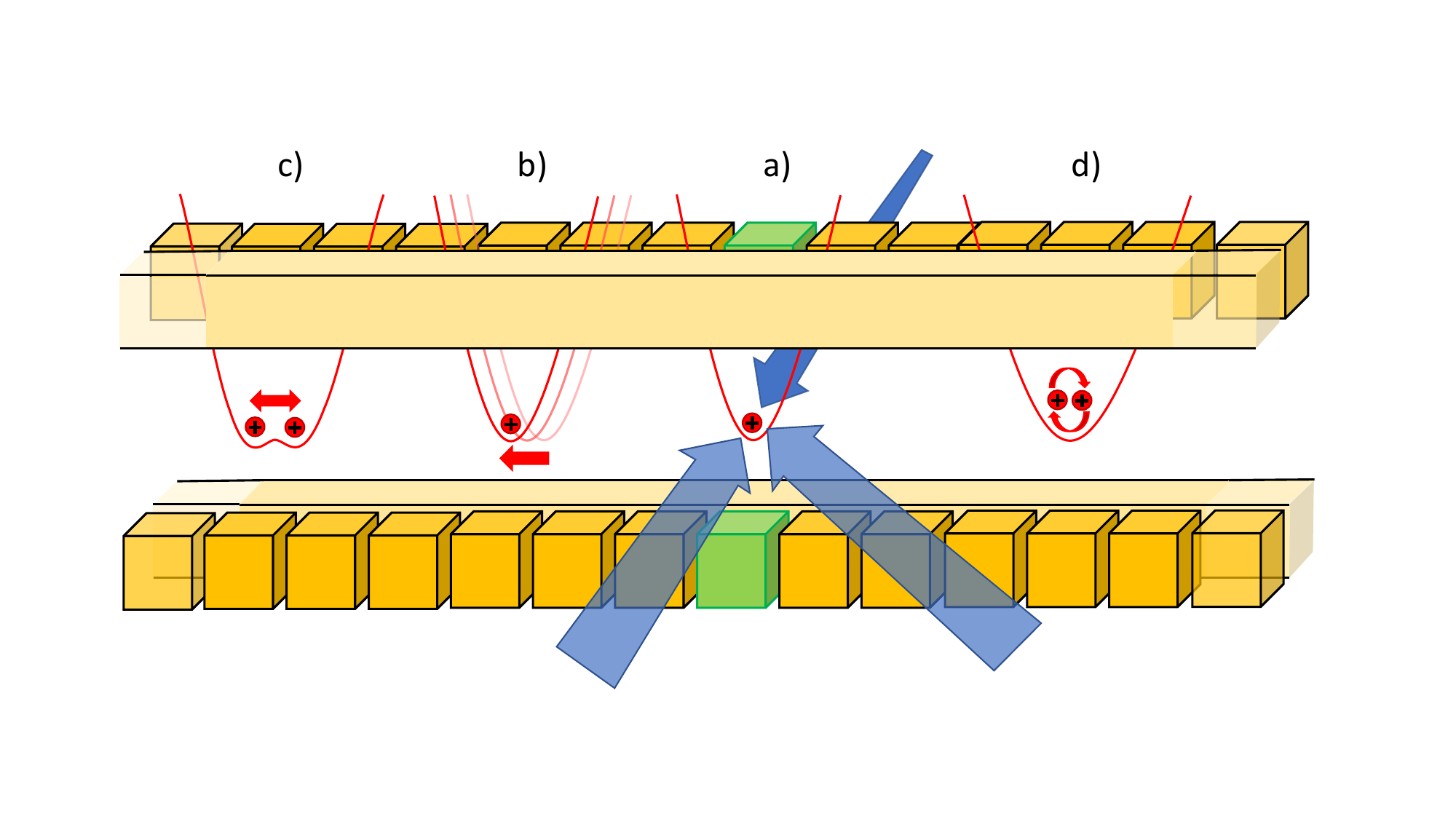}
\caption{Architecture of a linear segmented ion trap quantum computer along with operations. Trap segments are depicted in orange, the LIZ in green, and ions in red. Operations are: a) laser interaction in the LIZ, b) crystal displacement, c) crystal splitting, and d) crystal rotation.}
\label{fig:model}
    \vspace{-10pt}
\end{figure}

\section{Modeling of the architecture}
\label{Model}

As one concrete example, we use the architecture of the Mainz quantum computer, a linear segmented ion trap with 32 segments in which the LIZ is located at segment number~19 (Fig.~\ref{fig:model}). Even though the maximum number of ions in a processor is currently 98, we consider circuits with up to 200 to make the convergence and divergence clearly visible.
Such a large number also allows envisioning an expansion of the architecture. Furthermore, with such a maximum number of ions one can consider the linear segmented trap as a LIZ surrounded on both sides by a very large number of segments. 


To simulate this quantum computer's architecture,
or alternatively a sufficiently large ring storage region~\cite{helios}, a computer model was developed, which is based on an array of
objects (in the sense of object-oriented programming)
representing the segments. Each segment may contain a crystal that itself contains an ordered list of ions. An in-depth description of the computational data structure used has been described previously~\cite{1article}.

\begin{figure}[t]
    \centering
    \includegraphics[width=0.25\textwidth]{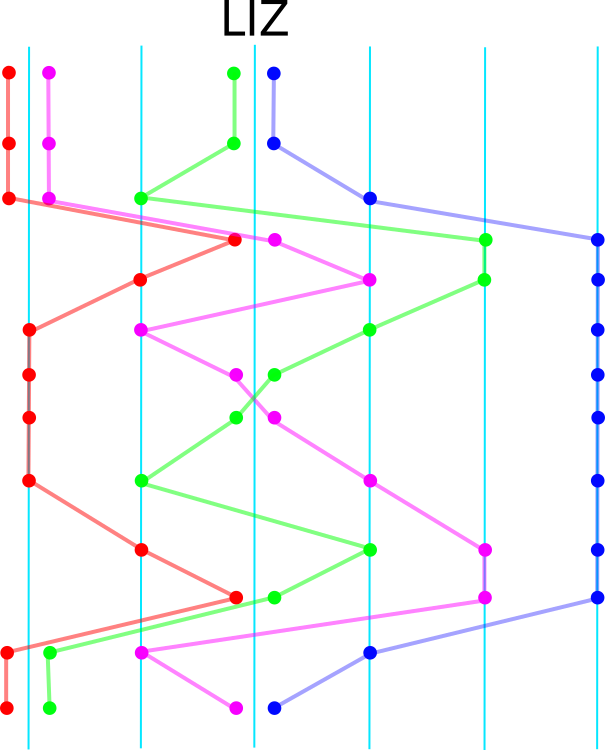}
    \caption{
    Illustration of an ion exchange. This figure can be either looked at from top to bottom or from bottom to top. The purple ion is exchanged with the green ion. A total of 3 merge and 3 split operations are involved.}
    \label{FigureCompilerCircuit}
    \vspace{-10pt}
\end{figure}

The costliest operations (merges/splits) are used in an ion exchange between two crystals. Fig.~\ref{FigureCompilerCircuit} illustrates an example. Such an exchange consists of splitting both crystals, forming a temporary crystal, rotating it (ion swapping), and splitting. Then, ions are merged to form two new crystals. Each ion exchange requires a total of 6 split and merge operations are required. This leads to an asymptotic value of 3 for an optimal circuit fit in the limit of a large number of qubits when only merges/splits are considered~\cite{1article}.

\section{Common ion order heuristic}
\label{sec:CIO}

The common ion order (CIO) heuristic proposed here is a greedy algorithm for the initial ordering of ion qubits. This algorithm is motivated by a mathematical proof for a class of quantum circuits having the same structure as the quantum Fourier transform (QFT). The proof showed that this circuit structure admits the existence of an optimal initial ion-qubit ordering for shuttling~\cite{1article}. This structure is composed of sub-circuits in which one ion is used in every gate of a sub-circuit. The heuristic identifies and uses this specific structure to find an efficient initial ordering 
leading to an optimal shuttling sequence. It must be mentioned that whether this optimal ordering can be found for any possible circuit, \textit{i.e.} including circuits that do not have a QFT-like structure, has yet to be proved.

Our heuristic for the algorithm is based on the concept of a common ion, that is an ion participating in a list of two-qubit gates that are contiguous in the circuit execution. The heuristic first focuses on  positioning such a common ion, with the rationale that it is algorithmically easier to optimize the shuttling of one ion  than several. Thus, the shuttling algorithm defines one ion as the common ion and seeks to optimize its position and path through the circuit.
The goal is to obtain an initial qubit ordering 
of the ions where the common ion is consistently one crystal away from the next ion involved in the following gate. Thus, for one exchange of ions, the common ion meets two different ions, one at the creation of the temporary crystal and a second when the exchange is done. This avoids useless exchanges, as the common ion takes part in a gate with every ion it meets.

The main objective of the algorithm is to place ions in such a way that their positions reflect their ordering in the gate list, where the common ion uses the shortest possible path to the next ion it must interact with. For this, each common ion is associated with a list of gates and with a list of ions ordered by their appearance in the list of gates.
The algorithm starts by scanning the circuit and identifying a sequence of common ions. The algorithm then associates to each common ion a sequence of gates involving the common ion. It then generates the placement of all ions involved in the circuit using these lists.

The input fed to the algorithm is the abstract circuit to be executed by the machine. In the following, a circuit of $N$ gates involving $M$ ions will be represented by a list $GL$ of all gates $g_i$ in the circuit, called the gate list, given by
\begin{equation}
  GL = [ g_0, g_1, \ldots, g_{N-1} ] .
\end{equation}
with $i$ being the order of the execution of the gate in the circuit.
Similarly,
the list of all ions, called the ion list will be given by
\begin{equation}
  IL = [I_0,I_1,\ldots,I_{M-1}] . 
\end{equation}
%

As the algorithm focuses on minimizing the number of splits and merges, one-qubit gates are not relevant as they do not influence the placement of ions. Hence, in describing algorithm, every gate $g_i$ can be considered as being a two-qubit gate unless said otherwise.

Concretely, the first part of the algorithm starts by identifying the common ion from the circuit input. For this, gates $g_0$ and $g_1$ are first fetched to find a common ion shared by these two gates. If $g_0$ and $g_1$ do not share an ion, then the algorithm selects the first ion participating in $g_0$ as the common ion with a gate list composed uniquely of $g_0$. Then the algorithm fetches the gates $g_1$ and $g_2$ to find a common ion. This process repeats until an ion shared by two contiguous gates is found.
Then, the algorithm iterates the gate list, adding them to the gate list associated with the common ion until it finds a gate $g_n$ that does not share the common ion. The algorithm then searches a common ion between $g_n$ and $g_{n+1}$. If there is no gate $g_{n+1}$, then the first ion of the gate $g_n$ is selected. 

Thus, the output of the first part of the algorithm is a list of common ions $Z$, each of these common ions $Z_i = I_i$ is associated with a list of gates $GL_i$ in which the ion is involved, this list having the form
\begin{equation*}
GL_i = [g_u,g_{u+1},\ldots,g_{u+v}]
\end{equation*}

The implementation of the algorithm to identify the common ions is given as a pseudo-code in Algorithm.\ref{AlIdentityCI}.

Once the common ions are identified and put in a list, all ions must be placed in the segmented trap. For this, the algorithm uses the gate list $GL_i$ associated with each common ion and the list $IL$ of all ions. The algorithm iterates over all two-qubit gates. For each gate, the non-common ion is extracted and if this ion is present in $IL$, it is added to the list of ions associated with the common ion and removed from $IL$. This is repeated for every common ion until $IL$ is empty. It must be noted that a common ion can be placed in a list of another common ion appearing sooner in the gate sequence.

\begin{algorithm}
\caption{Identifying common ions}
\label{AlIdentityCI}
\begin{algorithmic}[1]
   \STATE{$I$ := shared\_ions\_list($g_{0},g_{1}$)}
   \STATE{$Z$ := create\_new\_common\_ion\_list()}
   \STATE{$GL$ := get\_gate\_list()}
   
   \STATE{list\_of\_common\_ion\_list.append($Z$)}
   \IF{$I \ne \emptyset$}
    \STATE{$Z$.list\_common\_ion($I$)}
   \ELSE
    \STATE{$Z$.list\_common\_ion($g_{0}$.first\_ion())}
   \ENDIF

  \FORALL{$g_i \in G$}
   \IF{$g_i$.is\_ion\_in($GL_i$.get\_common\_ion())}
    \STATE{$Z$.associate\_gate\_to\_ion($g_i$)}
   \ELSE
   \STATE{$Z$ := create\_new\_common\_ion\_set()}
   \STATE{list\_of\_common\_ion\_set.append($Z$)}
   \STATE{$I$ := shared\_ions\_list($g_{i},g_{i+1}$)}
    \IF{$I \ne \emptyset$}
    \STATE{$Z$.set\_common\_ion($I$)}
    \ELSE
     \STATE{$Z$.set\_common\_ion($g_{i}$.first\_ion())}
    \ENDIF
    \STATE{$Z$.add\_gate($g_i$)}
   \ENDIF
  \ENDFOR
  \STATE{return list\_of\_common\_ion\_set}
\end{algorithmic}
\end{algorithm}

\begin{algorithm}
\caption{Ion positioning in the trap}
\begin{algorithmic}[1]
  \STATE $IL$ :=  ion\_list
  \STATE $V$ :=  empty\_list
  \FORALL{$Z \in$ list\_of\_common\_ion\_set}
 \FORALL{$g_i \in Z$.get\_gate\_list()}
 \STATE{ion := $Z$.get\_in\_common\_ion($g_i$)}
  \IF{ion $\in IL$}
   \STATE{$Z$.add\_to\_ions\_set(ion)}
   \STATE{$IL$.remove(ion)}
  \ENDIF
 \ENDFOR
 \STATE{$V$.place\_ion($Z$.get\_ions\_set()}
  \ENDFOR
\end{algorithmic}
\label{Al:PlacingIonSet}
\end{algorithm}

Once the list of common ions is created, the ions themselves must be placed in the segmented trap by generating an initial ion positioning list $V$. For this, the common ion lists $GL_i$ are processed by order of creation. Thus, the first common ion list is first placed in $V$, followed by the second common ion list placed in $V$, etc... This continues until all ions in the common ion list are placed. The pseudo-code in Algorithm~\ref{Al:PlacingIonSet} formalizes the previous discussion.

Algorithms~\ref{AlIdentityCI} and~\ref{Al:PlacingIonSet} called in that order form together the CIO algorithm. 


To generate a shuttling sequence, the program first extracts the list of gates from the input circuit provided in OPENQASM2.0 format~\cite{OPENQASM2.0}. The algorithm then uses this list to produce an object containing the common ions and the associated gates by using the CIO algorithm presented above. Then the shuttling algorithm proceeds to generate the shuttling sequence.
The shuttling algorithm has been presented in our previous article~\cite{1article}.

\section{Test methodology}
\label{sec:Imp}

\begin{figure*}[!t]
  \begin{center}
    \begin{tabular}{ccc}
      \includegraphics[width=0.3\linewidth]{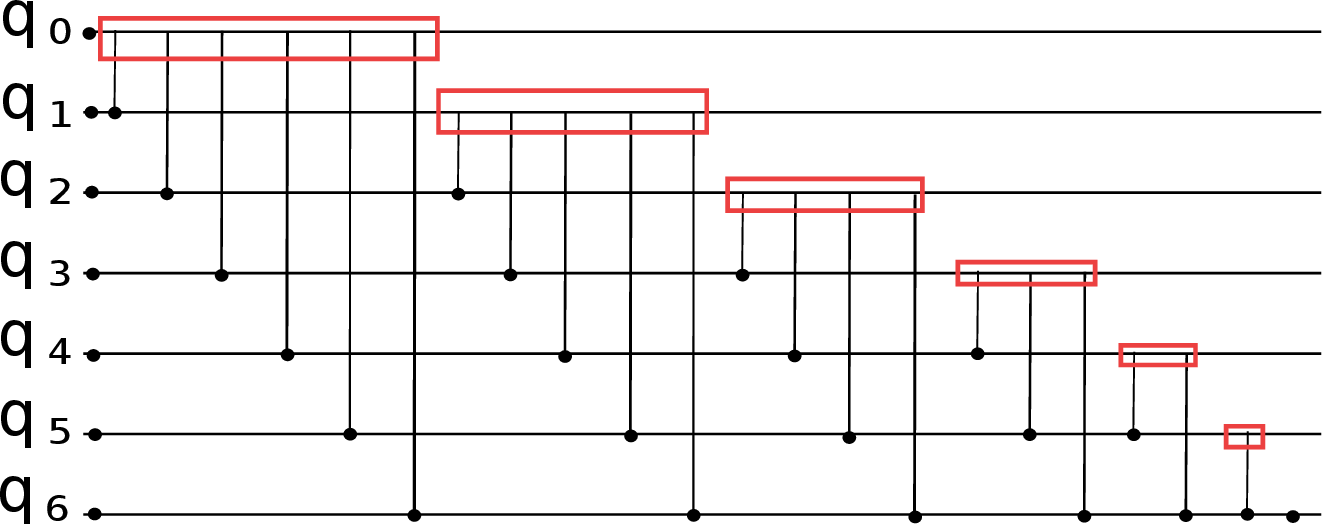} &
      \includegraphics[width=0.3\linewidth]{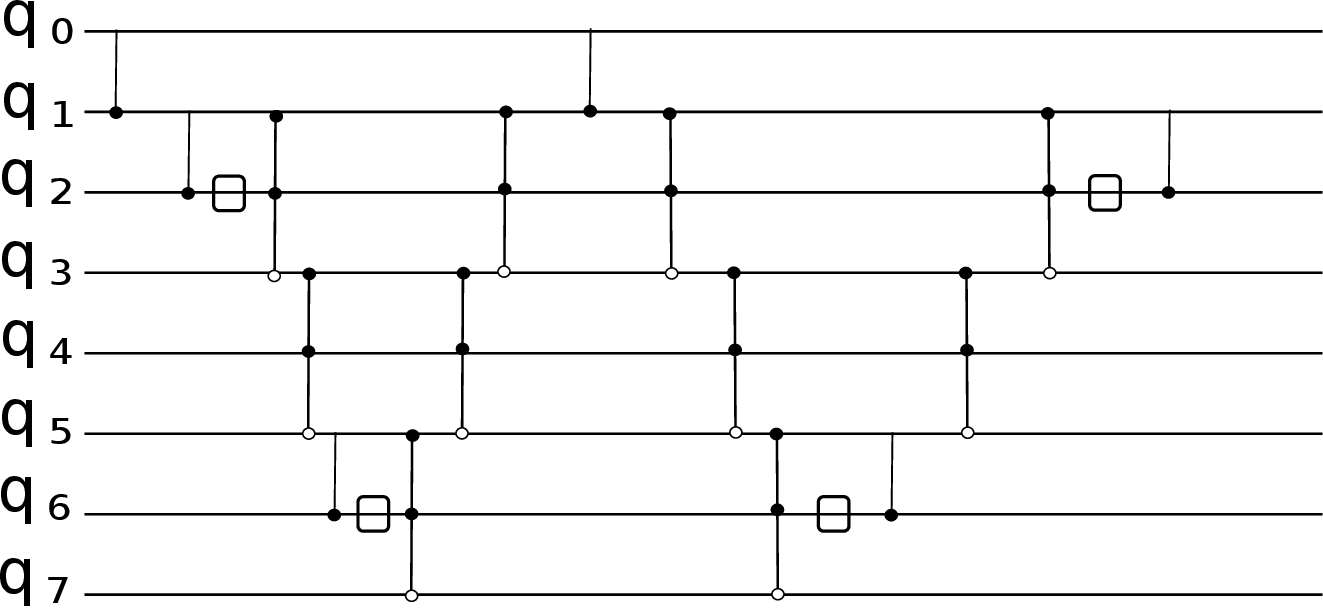} &
      \includegraphics[width=0.3\linewidth]{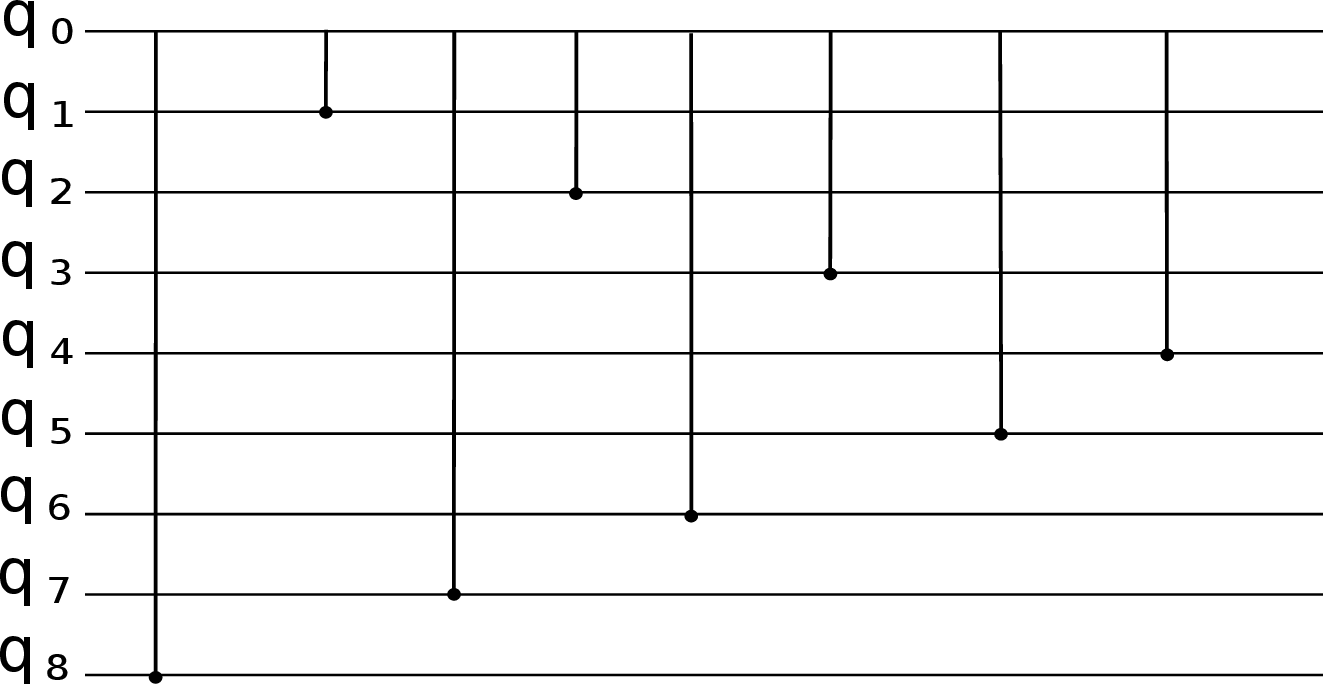}
       \\
      (a) QFT & (b) Carry & (c) Yoyo\\
      \includegraphics[width=0.3\linewidth]{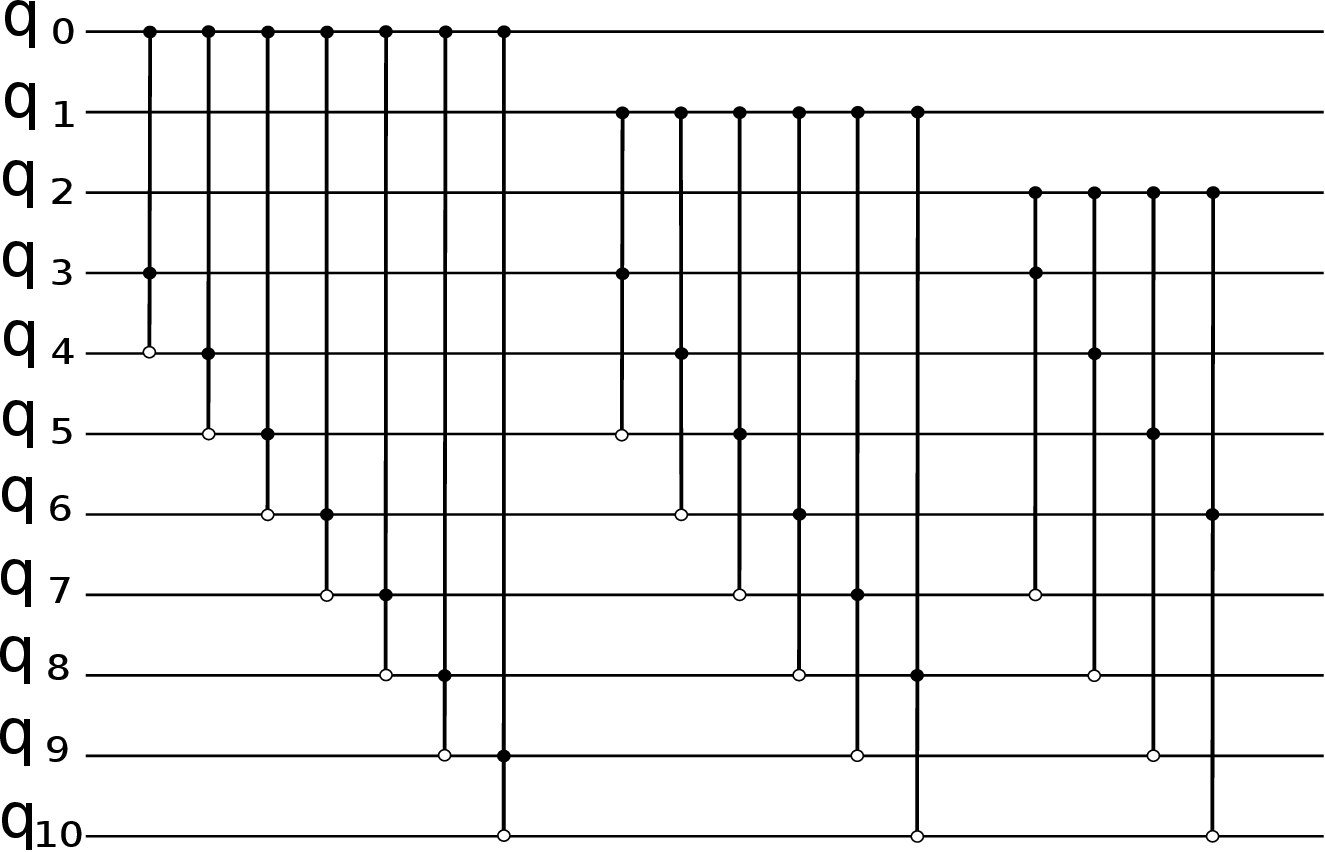} &
      \includegraphics[width=0.3\linewidth]{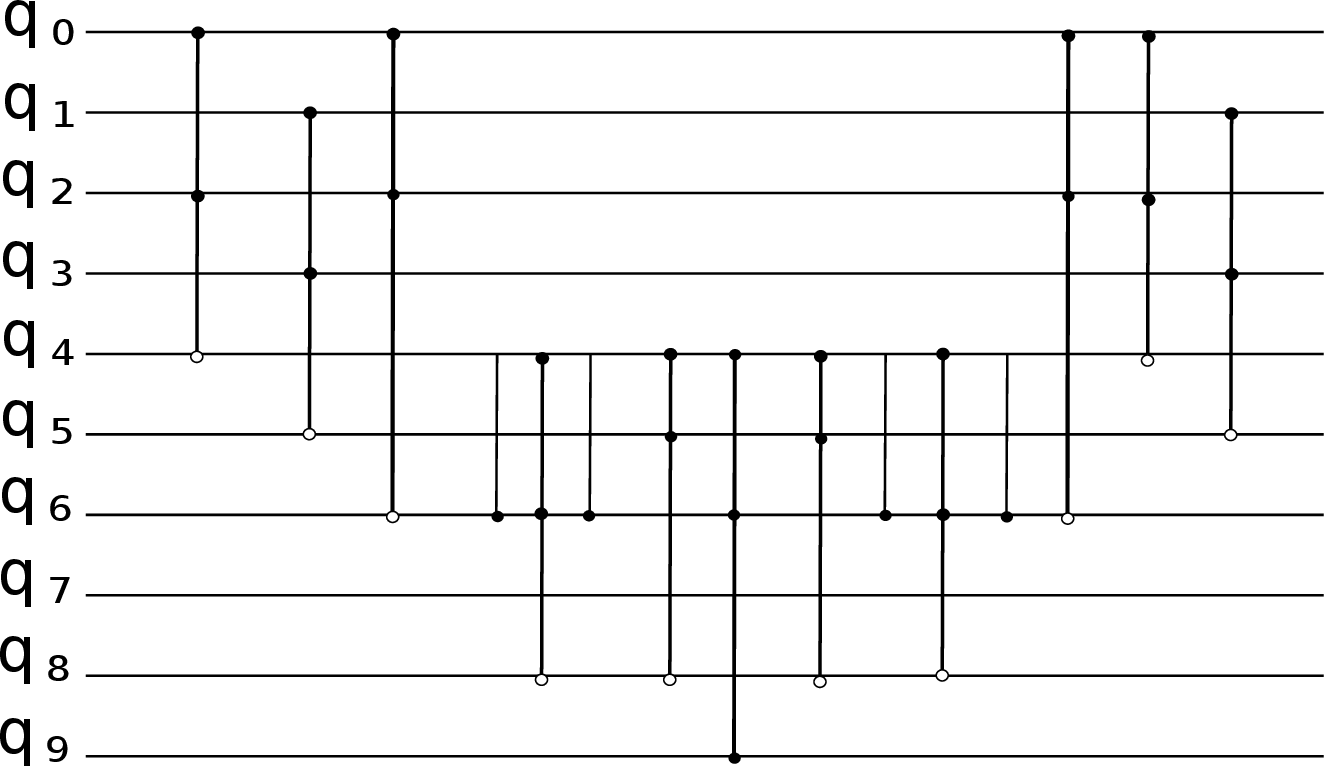} &
      \includegraphics[width=0.3\linewidth]{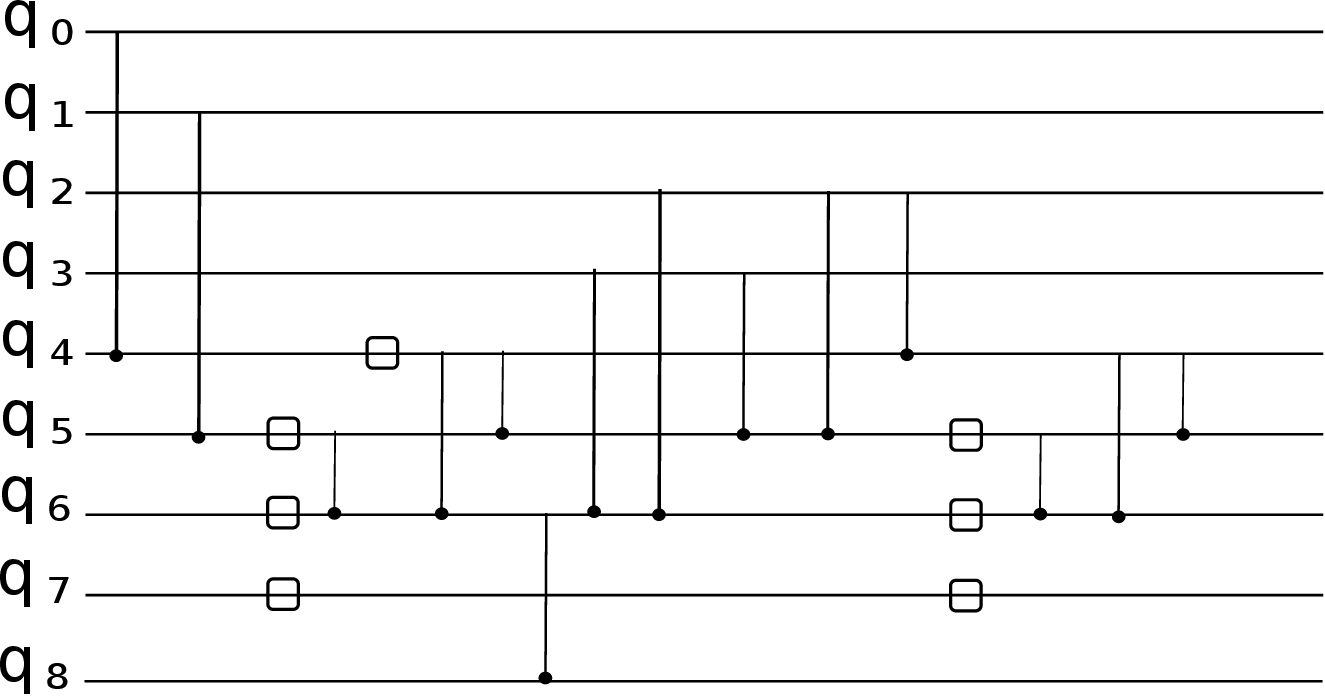} \\
      (d) Shift & (e) Comparator & (f) Adder
    \end{tabular}
  \end{center}
  \caption{Circuits used for testing the algorithm proposed herein. Squares represent one-qubit gates, while in Toffoli gates the filled circles correspond to the control qubits and the empty circles to the controlled qubits. The number of qubits in the circuits shown differ from one circuit to another; these numbers are not important, they were chosen just for the purpose of making the circuit structures clearly visible. Furthermore, we added red rectangle to QFT to indicate common ion as example. Note that common ions are highlighted in red in~(a), since the QFT circuit easily lends itself for such highlighting.}
  \label{fig:CircuitDesc}
\end{figure*}

To generate the results presented in next section, two
initial ion-qubit ordering 
algorithms
are applied on different circuits. The first is the CIO algorithm described above. The second is the \textit{order as is} (OAI) algorithm, which serves as a benchmark.
The OAI was published in our previous work~\cite{1article}, and provided competitive performances compared to the other approches then published, hence its choice to serve as a benchmark here.
The OAI simply assigns qubit indices
left-to-right from~1 to~$N$ according to the initial ion positions in the segmented trap. Each circuit is tested for an increasing number of ions from 2 to 200 with the step between consecutive numbers depending on the type of circuit as not all circuits can be expanded by the addition of one ion.
The circuits considered are: QFT, Carry, Yoyo, Shift, Adder, and Comparator; they will now be briefly described.

      

The QFT circuit~\cite{QFT} (Fig.~\ref{fig:CircuitDesc}~(a)) is well-known. Its structure is ideally suited for the CIO algorithm; it has in fact served as the initial motivation for developing the  CIO algorithm. The QFT is an optimal circuit for a linear architecture, see \cite{1article}.

The Carry circuit~\cite{Carry} (Fig.~\ref{fig:CircuitDesc}~(b)) is frequently used in classical computing. It is thus interesting to study it in the context of quantum computing. The Carry circuit uses the Toffoli gate, which appears in a large number of algorithms and is notoriously expensive to implement. The Carry circuit allows testing the algorithm in the presence of this gate.


The Yoyo and Shift circuits (Figs.~\ref{fig:CircuitDesc}~(c) and~(d)) have been devised to test specific aspects of the algorithm.
The Yoyo circuit allows studying the behavior of the algorithm with a circuit structure that requires ions far from each other to interact.
Thus, if the algorithm functions properly, the Yoyo circuit should lead to a circuit fit that tends asymptotically to the optimal value of 3 as every ion exchange allows for two gates to be executed.
The Shift circuit comprises exclusively Toffoli gates. As depicted, three qubits interact with every other qubit through a Toffoli gate. This circuit was selected to investigate the behavior of the algorithm in presence of Toffoli gates.
Finally, the Comparator and Adder circuits (Figs.~\ref{fig:CircuitDesc}~(e) and~(f)) have been chosen for their complexity since, as can be seen, they are composed of sub-circuits such as the QFT and the Toffoli gates, thus allowing to see the behavior of the CIO with such composed circuits.

\section{Results for the CIO algorithm}
\label{sec:ResultsCIO}

\begin{figure*}[htbp]
\centering
    \subfloat{
      \includegraphics[width=0.28\linewidth]{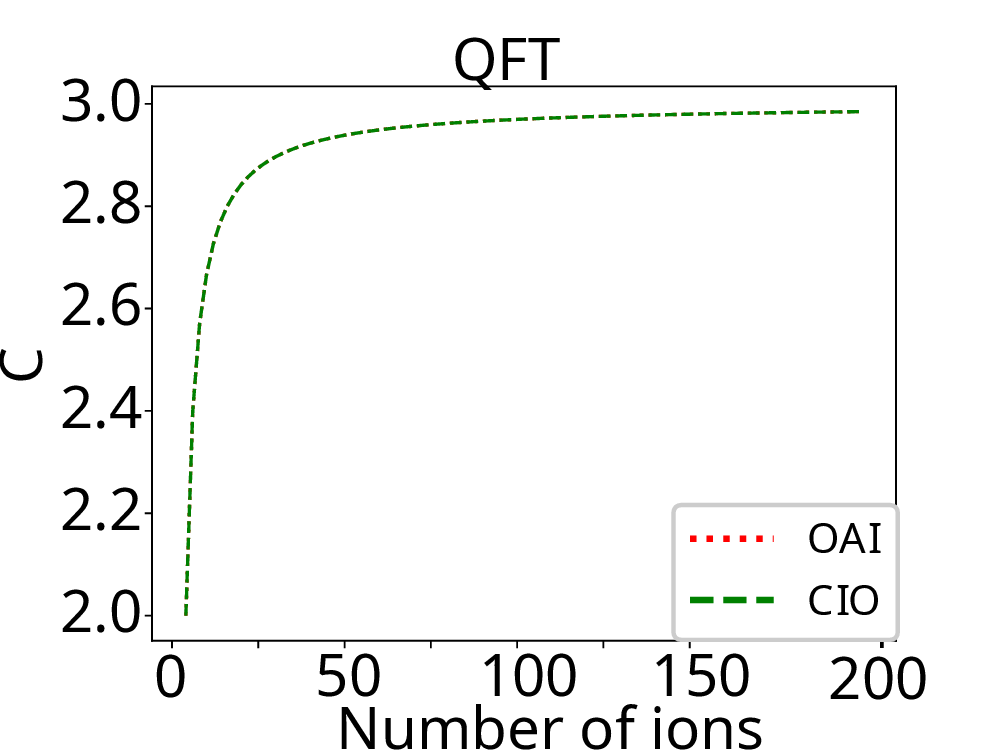}
      \label{subfig:QFTFit}
    }
    \subfloat{
      \includegraphics[width=0.28\linewidth]{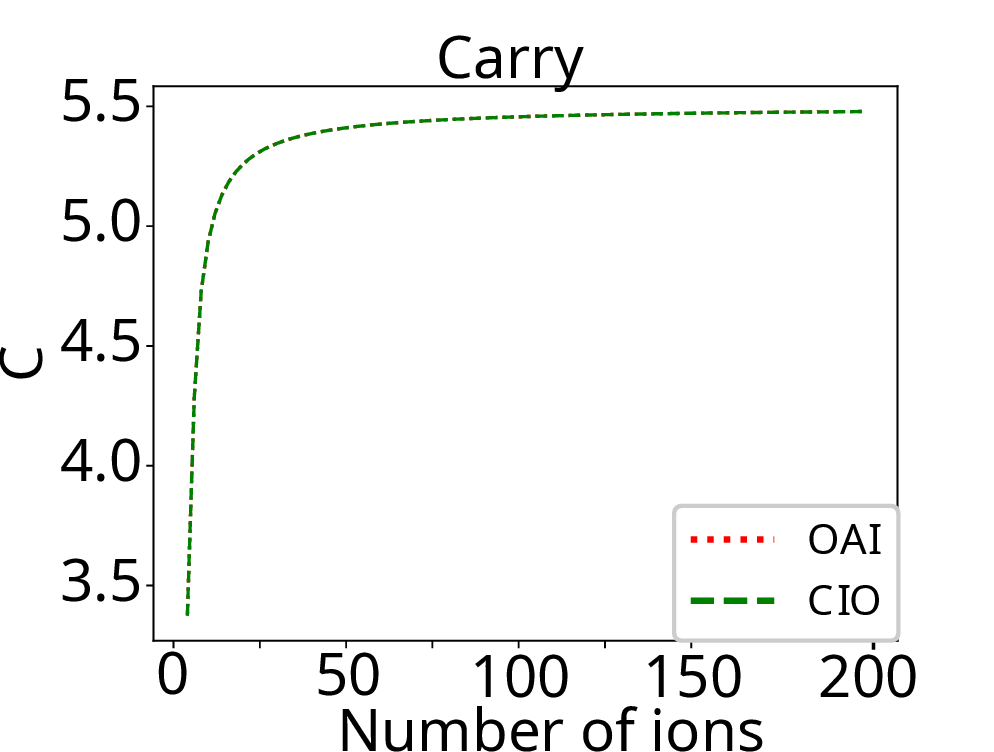}
      \label{subfig:CarryFit}
    }
    \subfloat{
      \includegraphics[width=0.28\linewidth]{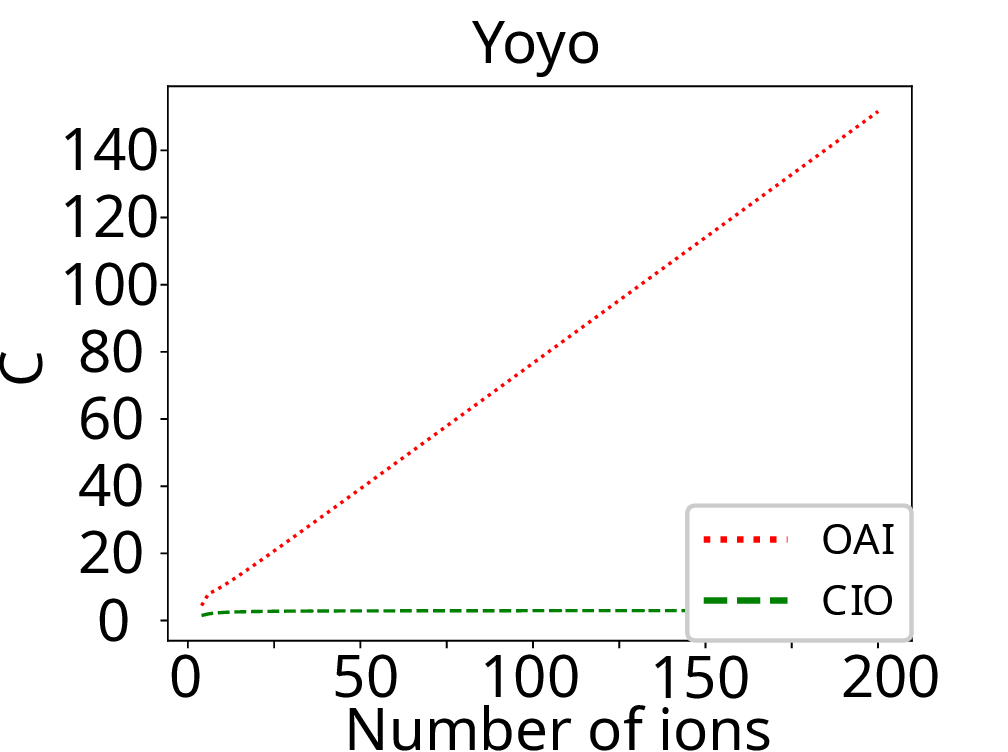}
      \label{subfig:YoyoFit}
    } \\
    \subfloat{
      \includegraphics[width=0.28\linewidth]{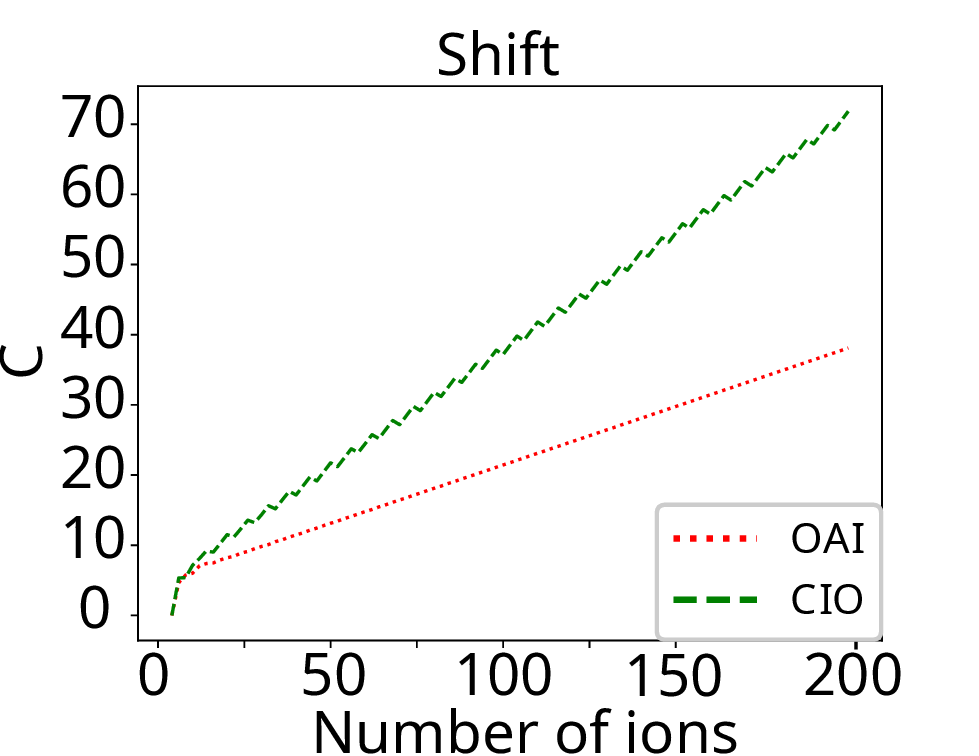}
      \label{subfig:ShiftFit}
    }
    \subfloat{
      \includegraphics[width=0.28\linewidth]{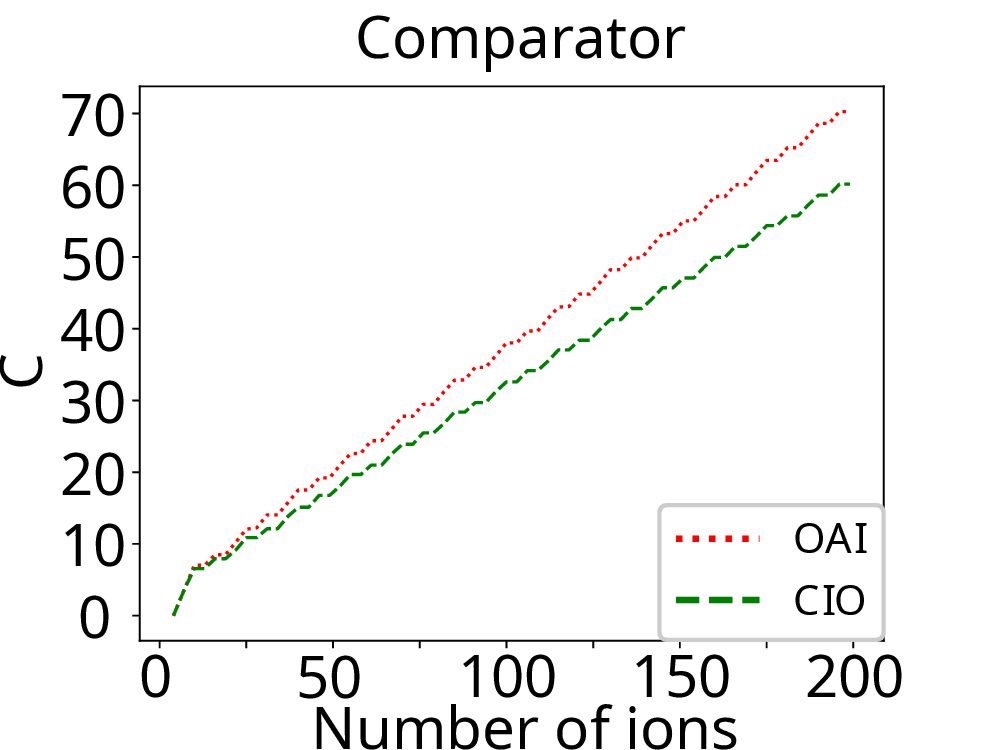}
      \label{subfig:ComparatorFit}
    }
    \subfloat{
 \includegraphics[width=0.28\linewidth]{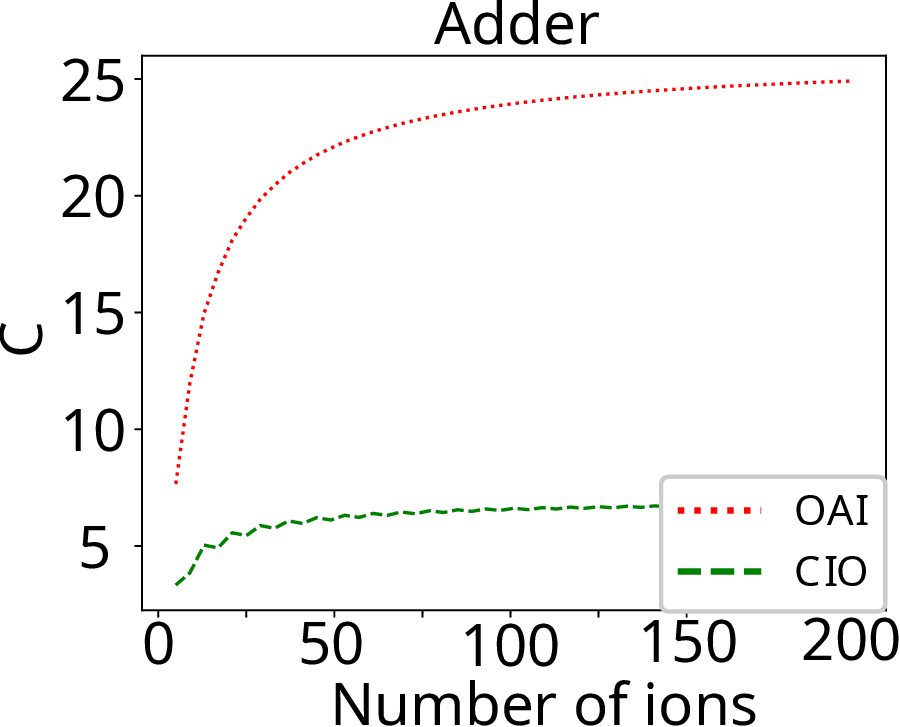}
 \label{subfig:AdderFit}
}
\caption{Results comparing the CIO and OAI initial ordering algorithms in terms of the circuit fit metric for the circuits considered. Note that for the QFT and Carry, the red and green curves overlap, hence the red curves not being visible.
}
\label{fig:Result}
\end{figure*}

\begin{figure*}[htbp]
\centering
\subfloat{
\includegraphics[width=0.28\linewidth]{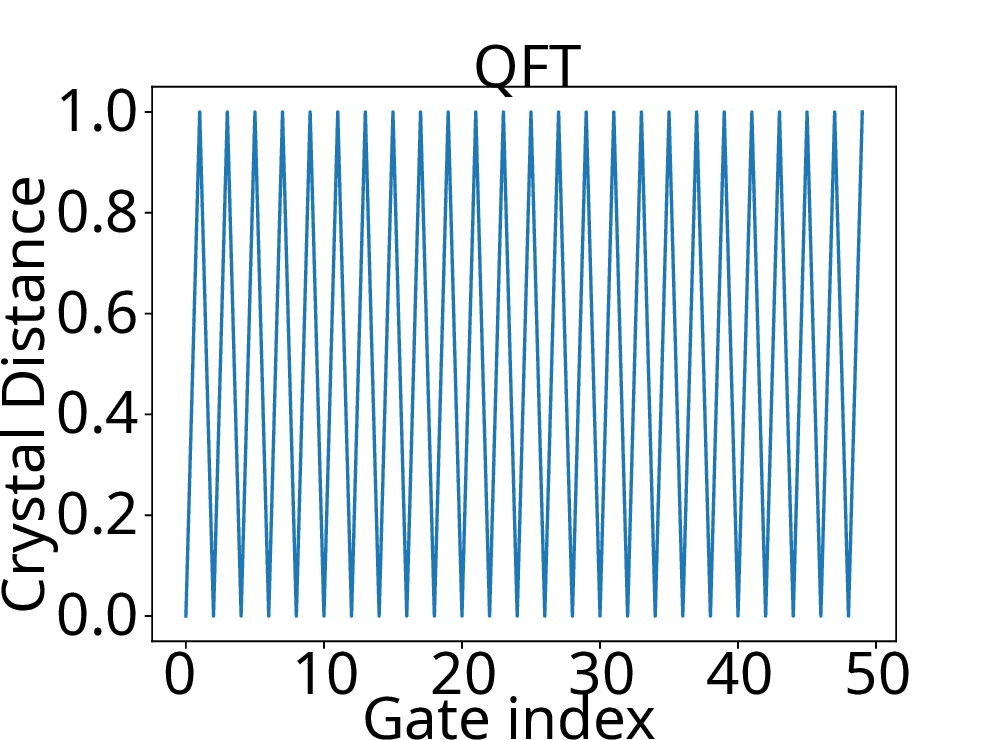}
\label{subfig:QFTInside}
}
\subfloat{
\includegraphics[width=0.28\linewidth]{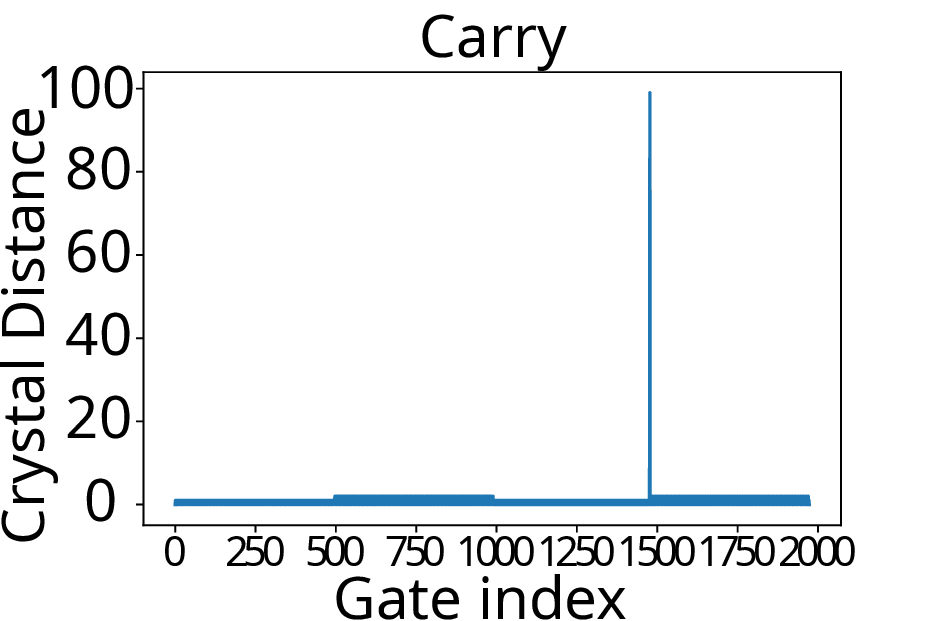}
\label{subfig:CarryInside}
}
\subfloat{
\includegraphics[width=0.28\linewidth]{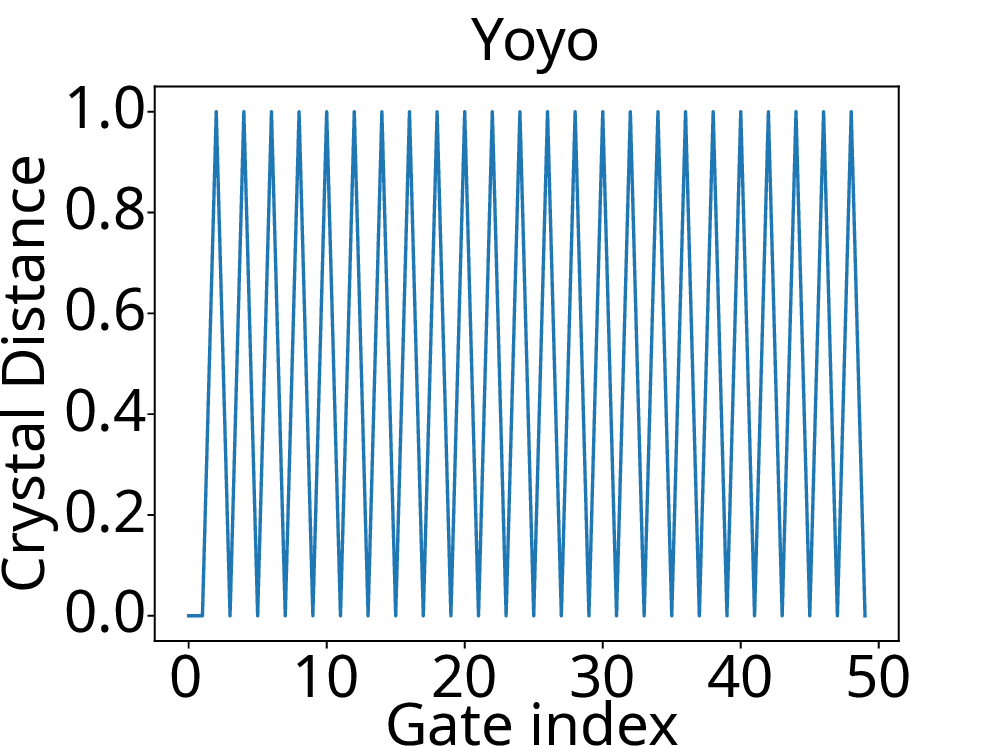}
\label{subfig:YoyoInside}
}\\
\subfloat{
\includegraphics[width=0.28\linewidth]{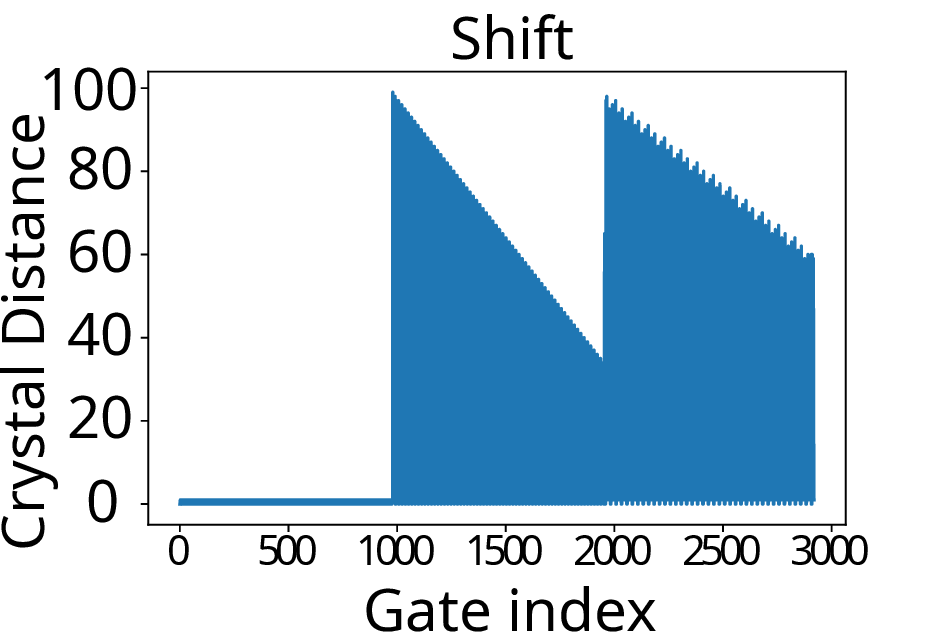}
\label{subfig:ShiftInside}
}
\subfloat{
\includegraphics[width=0.28\linewidth]{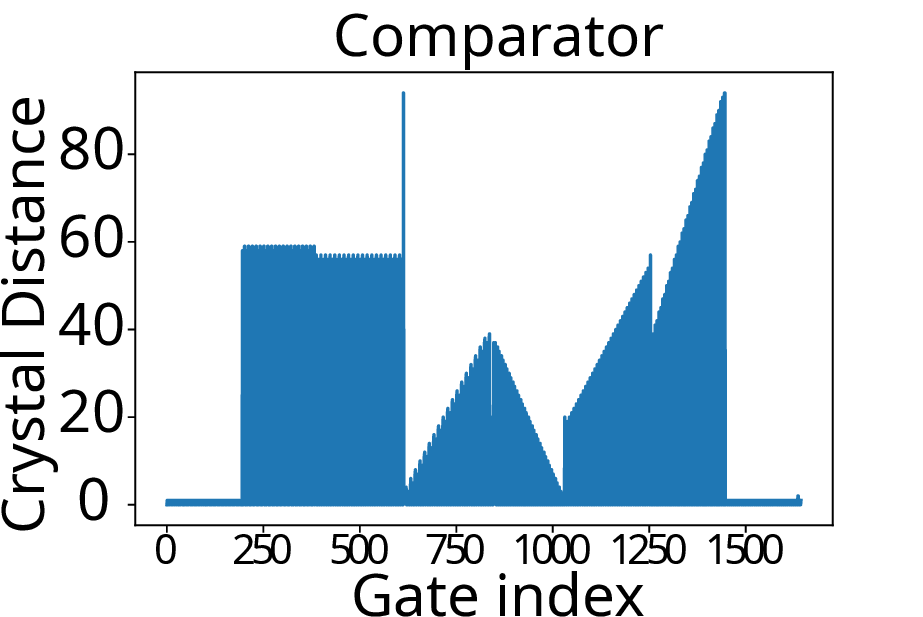}
\label{subfig:ComparatorInside}
}
\subfloat{
\includegraphics[width=0.28\linewidth]{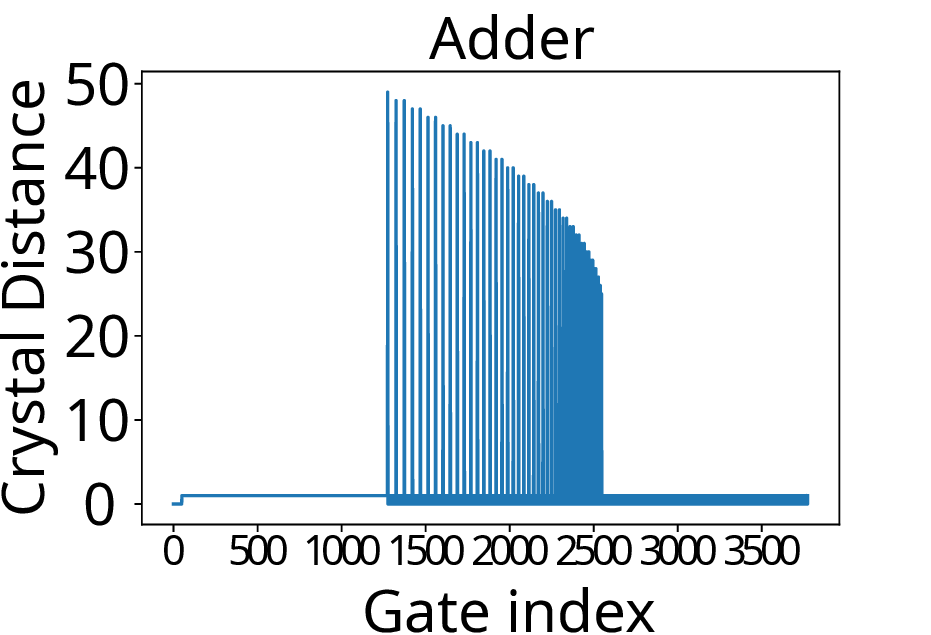}
\label{subfig:AdderInside}
}

\caption{Evolution of the crystal distance between ions participating in two-qubit gates while shuttling proceeds.
Note that, for clarity, the graphs for the QFT and Yoyo circuits only show the first 50 gates since  the oscillatory behavior is the same for the entire graph up to their full number of gates.
}
    \label{fig:ResultInside}
\end{figure*}

Fig.~\ref{fig:Result} compares the CIO and OAI algorithms in terms of the circuit fit, thus the effort in split and merge qubit register operations, for the six circuits presented previously.

\begin{itemize}
    \item[(a)] QFT: The circuit fit is the same for the CIO and OAI. This is consistent with the fact that the OAI is the optimal placement for the QFT in a linear architecture.

    \item[(b)] Carry: The CIO and OAI have equivalent behaviors indicating that the Carry circuit possesses a QFT-like structure. 
        
    \item[(c)] Yoyo: The CIO is efficient as it converges to the optimal circuit fit while the OAI diverges.
    
    \item[(d)] Shift: The CIO has a worse behavior than the OAI. This is an instance showing that the CIO is not well adapted to Toffoli gates, which will be explained later, as they lead to divergent results.

    \item[(e)] Comparator: The CIO has a better behavior than the OAI, although both diverge. 

    \item[(f)] Adder: The circuit fit for the CIO is much better than for the OAI.
    
\end{itemize}



These results show that the CIO has better or equal performance in terms of circuit fit when compared to OAI for most circuits. This shows the interest of CIO algorithm. The fact that for one circuit (Shift), the CIO gives worse results than the OAI indicates that the CIO is not best suited for circuits with Toffoli gates.
It can be concluded from this that the CIO is an efficient algorithm for the initial
ion-qubit ordering 
prior to shuttling. There is, however, still room for improvement.

Another means of studying the behavior of the algorithm in terms of circuits is to follow during shuttling the evolution of the distance between the two ions that must take part in a two-qubit gate at the point this gate is to be executed in the circuit.
This distance, to be called the \textit{crystal distance} (CD), is defined as the number of crystals contained in the segments between the two ions taking part in the gate prior to the shuttling required
to execute this particular gate.
The CD is then plotted against the indices of the gates in the circuit, as the gates are executed one after the other. The rationale for following such a distance is that the lower it is maintained by the algorithm, the less shuttling operations are required, and hence the less costly shuttling will be.

For each circuit, a number of two hundred ions is considered, or as close as possible to 200 depending on the circuit. Only two-qubit gates are considered in the evaluation of the CD. It may happen in rare instances that two consecutive two-qubit gates targeting the same ions can be executed together without any ion exchange. Such a rare event may cause a difference between the number of gates shown and the total number of two-qubit gates in the circuit. 
Results for the CIO algorithm are presented in Fig.~\ref{fig:ResultInside}. The CD parameter shows the evolution of the disorganization of the ions placement with respect to the two-qubit gates to be executed in the circuit as shuttling is carried out.
The results show that most of the circuits can be partitioned into sub-circuits, each with a specific trend for the evolution of the crystal distance.

The results for the QFT and Yoyo circuits in Fig.~\ref{fig:ResultInside}~(a) and~(c) show a perfect optimization with no ions separated more than one crystal away.
The Carry circuit in Fig.~\ref{fig:ResultInside}~(b) shows a stable distance until a peak around gate 1500. This peak corresponds to one qubit being pushed out to one end of the chain of crystals until it is needed. 
The results in Fig.~\ref{fig:ResultInside}~(d) and~(e) show that  there is a range of gate indices (horizontal axis) for the Adder and Comparator circuits in which the shuttling is not optimal. As a whole, Fig.~\ref{fig:ResultInside} shows that the CIO algorithm is highly efficient for simple circuits, but has difficulties with circuits composed of different sub-circuits as the ion positioning of one sub-circuit is not best suited to begin the next sub-circuit. 

Fig.~\ref{fig:ResultInside} also shows that circuits tend to have sub-circuits in which the distance stays small.  Although the performance for the Carry and Adder circuits is not as affected as for other circuits, the results and the structure of these circuits show why the optimal circuit fit is not reached for them. It is seen that each sub-circuit for the Comparator and Shift circuits tends not to give an efficient arrangement of ions for the shuttling in the next sub-circuit that follows it.  It is also observed that sub-circuits tend to converge for the Shift and Adder circuits, \textit{i.e.} the shuttling cost for the sub-circuits decreases as the arrangement of the circuit tends to get closer to the optimal placement. It is seen as well that the Comparator circuit is divergent, \textit{i.e.} the crystal distance increases with the gate index, while the Shift circuit is convergent. This
shows 
that the initial
ion-qubit
ordering has a large influence on the shuttling.
\begin{figure}[t]
\centering
\includegraphics[width=0.30\textwidth]{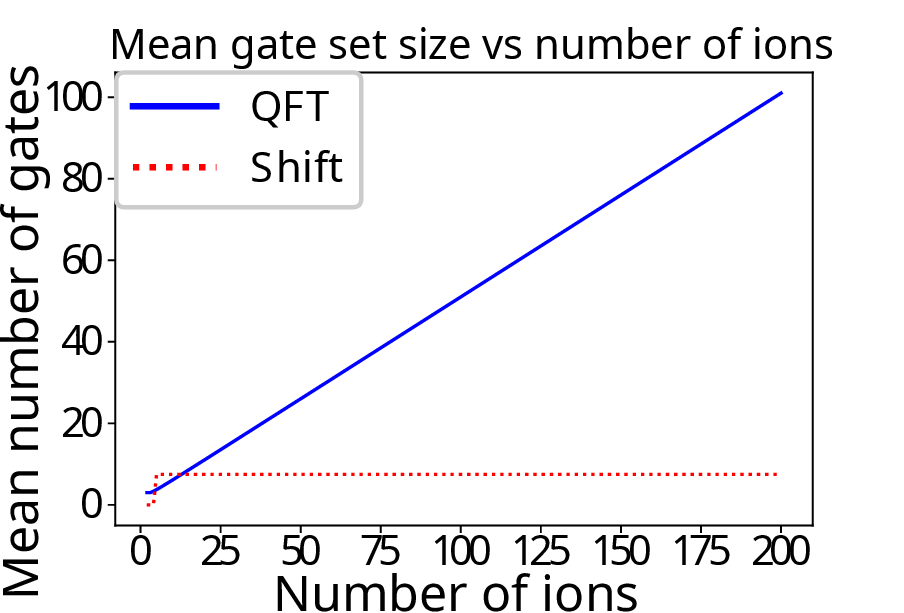}
\caption{Mean number of gates per common ion versus the number of ions.}
\label{fig:RatioMI}
\end{figure}

The Toffoli gate causes difficulties as it involves three interacting qubits. This breaks the common ion logic as there is no common ion inside the Toffoli gate itself but rather a block of common ions. Thus, even if the common block in the Shift circuit can visually be seen, the CIO algorithm does not see this block. This is illustrated for the QFT and Shift circuits in Fig.~\ref{fig:RatioMI}, which compares the mean number of gates per common ion list as a function of the number of ions. 

For the Shift circuit, the mean number of gates per common ion tends to 7.5, whereas for the QFT, this mean number grows  linearly with the number of ions. Other results not presented here show that the mean number of gates per common ion also tends to 7.5. For circuits involving Toffoli gates, it is seen the CIO algorithm fails to create large ensembles of gates sharing a common ion as every Toffoli gate creates its own list of ions.
One way to resolve this issue would be for the algorithm to shuttle blocks of ions. Identifying such blocks appears be an algorithmic problem of great complexity, and, furthermore, the shuttling of these blocks itself comes with a large cost. The use of more ions per crystal could perhaps resolve the shuttling cost issue, but the physical difficulty to rotate large crystals makes this a theoretical solution at best for now. 

\section{Ion reorganization}
\label{sec:IonReorg}

\begin{figure*}[htbp]
\centering
    \subfloat{
      \includegraphics[width=0.3\linewidth]{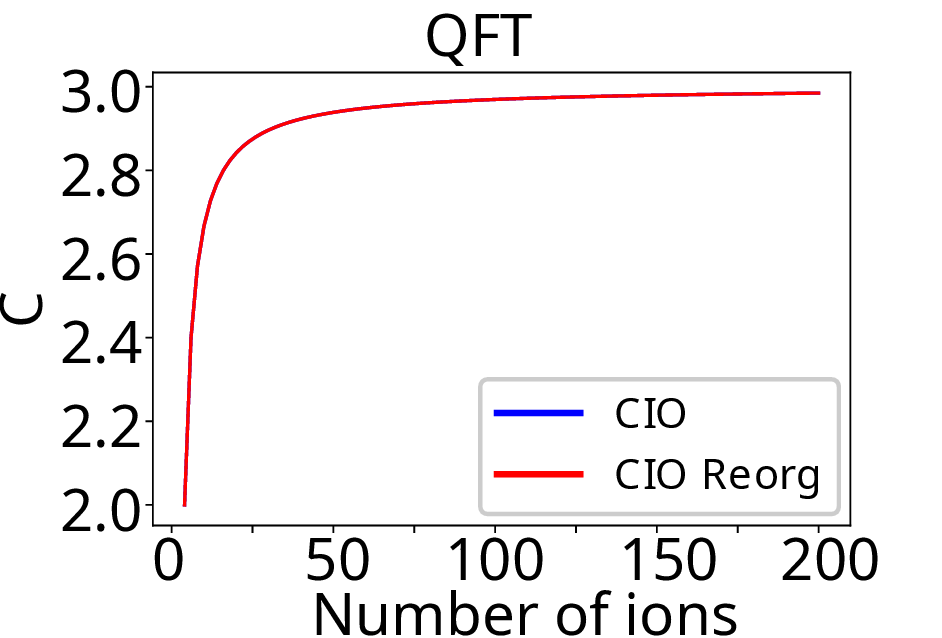}
      \label{subfig:QFTReorgGraph}
    } 
    \subfloat{
      \includegraphics[width=0.3\linewidth]{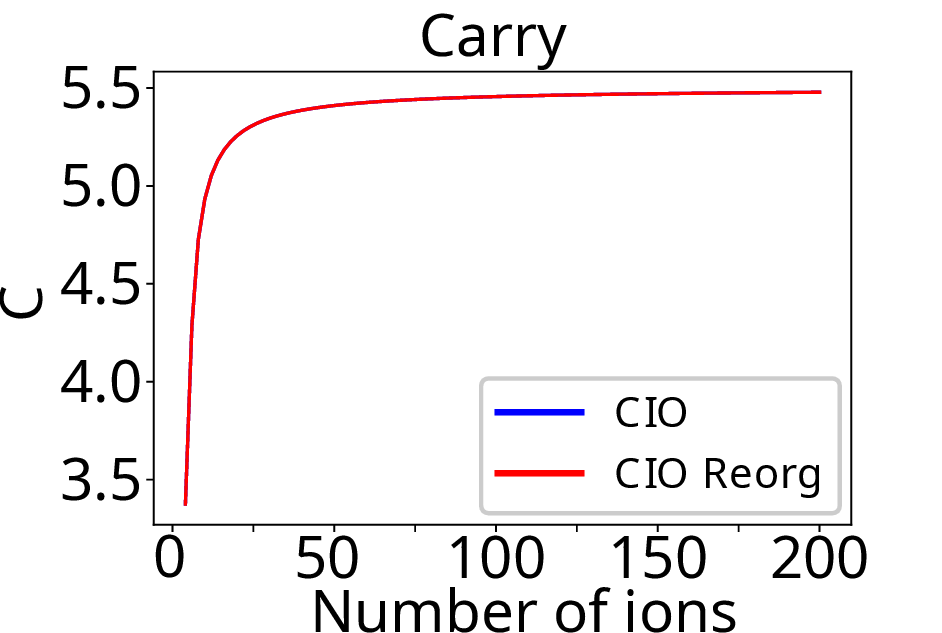}
      \label{subfig:CarryReorgGraph}
    } 
    \subfloat{
      \includegraphics[width=0.3\linewidth]{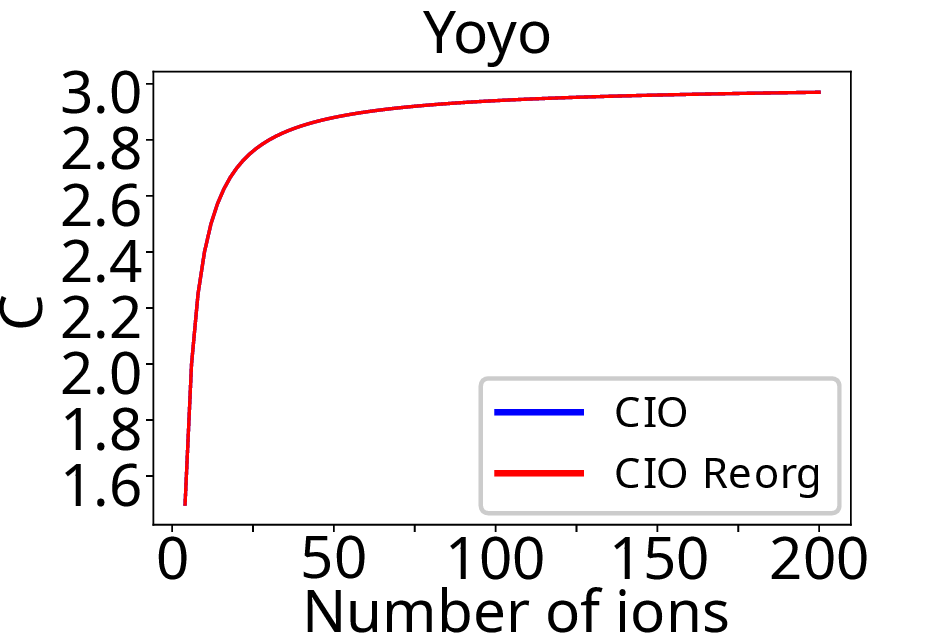}
      \label{subfig:YoyoReorgGraph}
    } \\
    \subfloat{
      \includegraphics[width=0.3\linewidth]{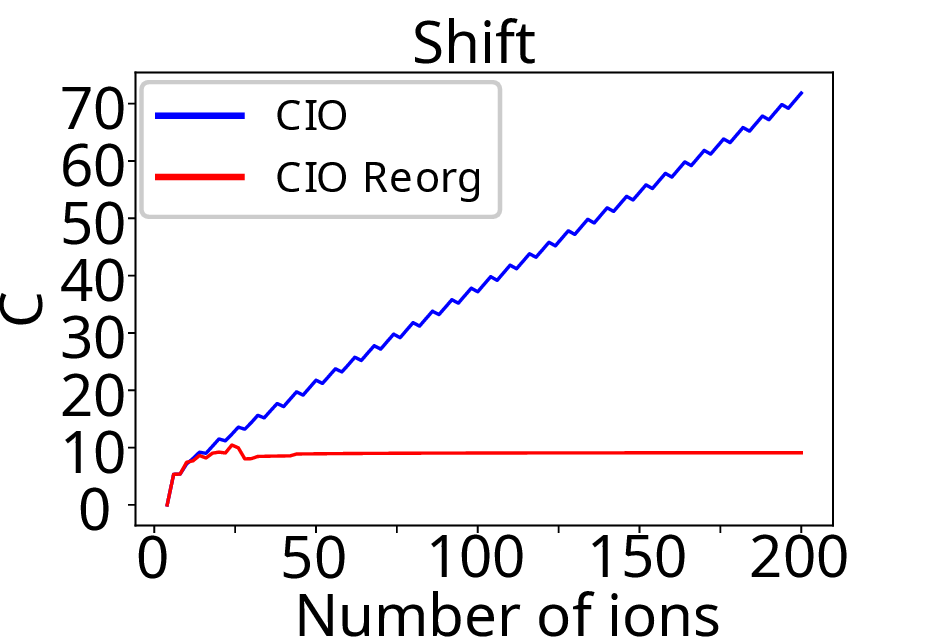}
      \label{subfig:ShiftReorgGraph}
    }
    \subfloat{
      \includegraphics[width=0.3\linewidth]{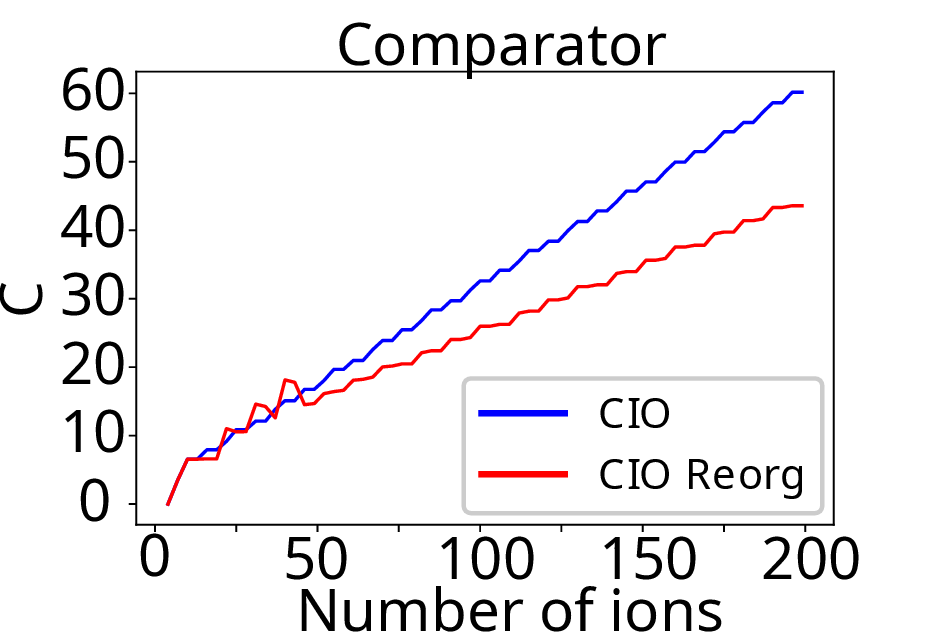}
      \label{subfig:ComparatorReorgGraph}
    } 
    \subfloat{
      \includegraphics[width=0.3\linewidth]{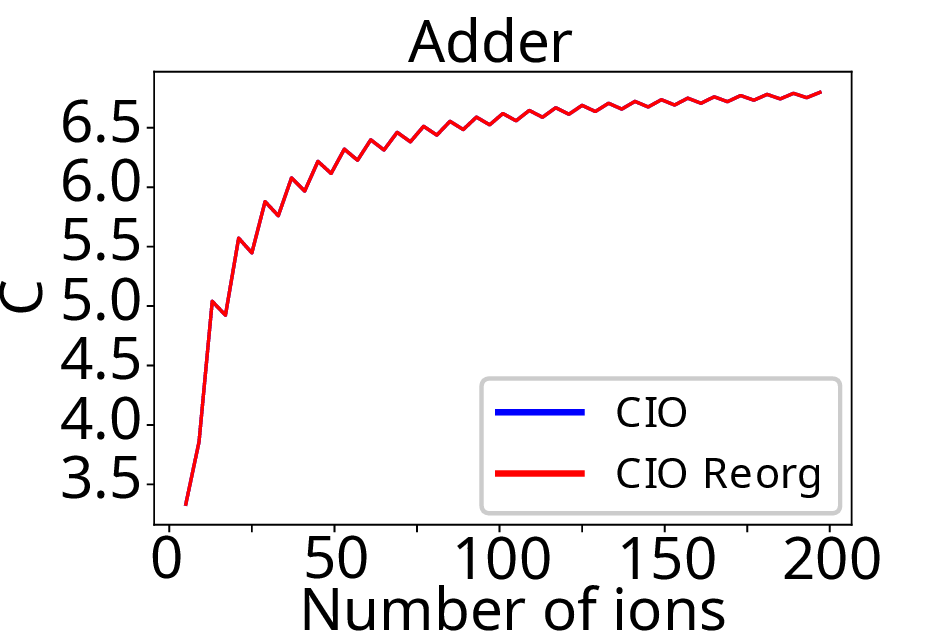}
      \label{subfig:AdderReorgGraph}
    }
	
    \caption{Circuit fit ($C$) results for the reorganization: CIO 
    in blue and CIO with reorganization in red. Note: For the QFT, Carry, and Adder, the red and blue curves overlap, hence the blue curves not being visible.
}
    \label{fig:ResultReorg}
\end{figure*}

\begin{figure*}[htbp]
\centering
    \subfloat{
      \includegraphics[width=0.3\linewidth]{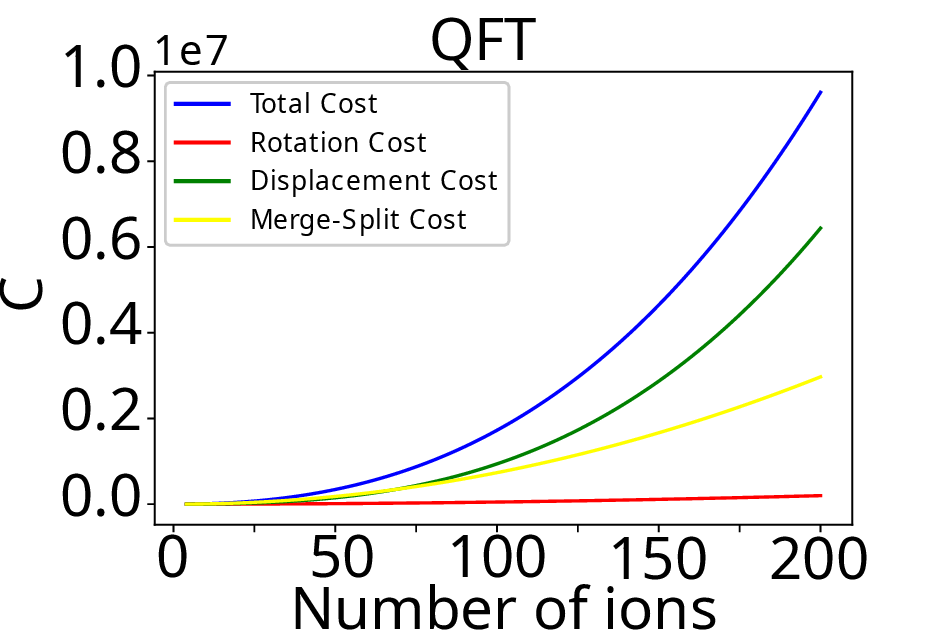}
      \label{subfig:QFTTotGraph}
    } 
    \subfloat{
      \includegraphics[width=0.3\linewidth]{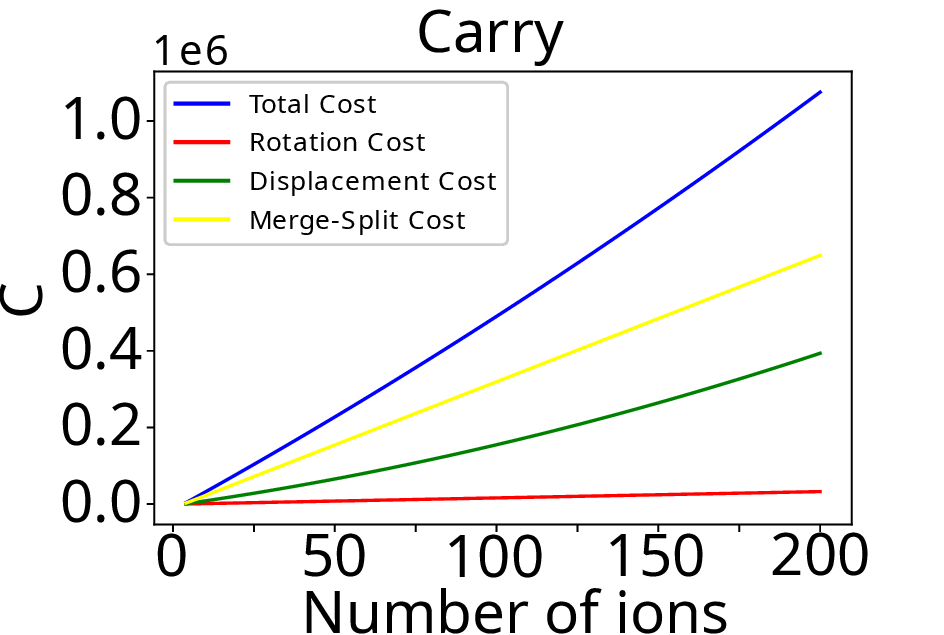}
      \label{subfig:CarryTotGraph}
    }
    \subfloat{
      \includegraphics[width=0.3\linewidth]{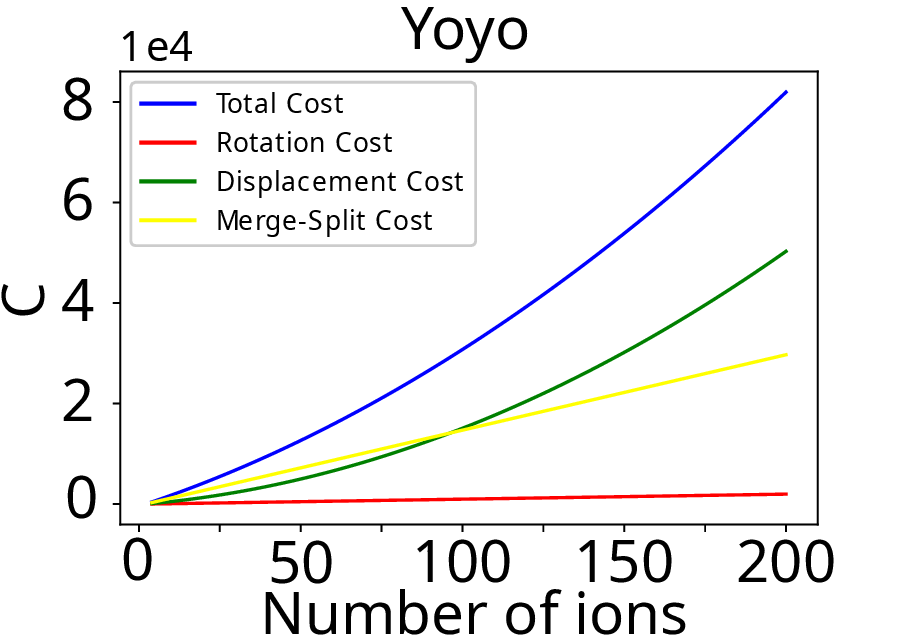}
      \label{subfig:YoyoTotGraph}
    } \\
    \subfloat{
      \includegraphics[width=0.3\linewidth]{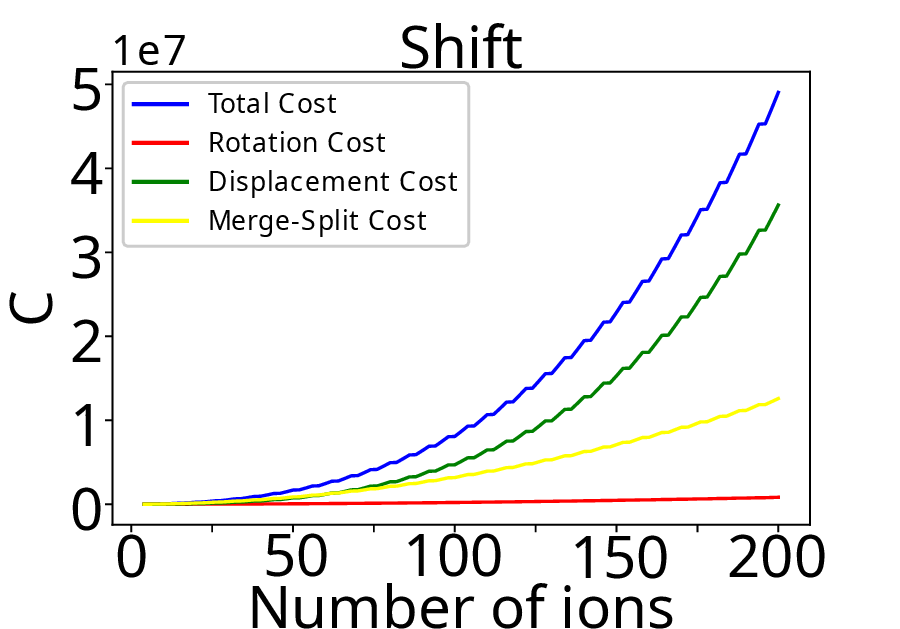}
      \label{subfig:AdderTotGraph}
    }
    \subfloat{
      \includegraphics[width=0.3\linewidth]{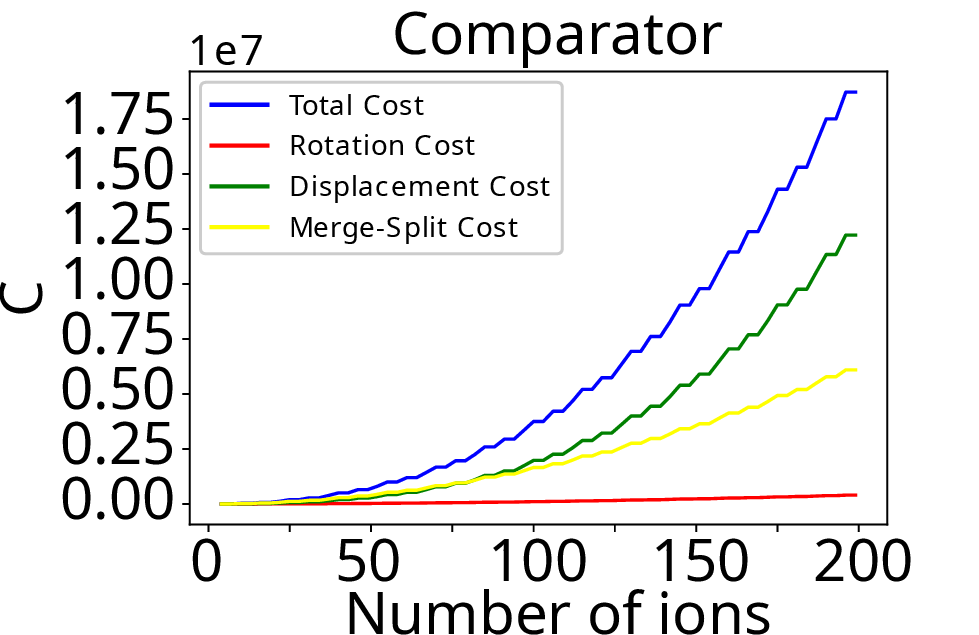}
      \label{subfig:ComparatorTotGraph}
    }
    \subfloat{
      \includegraphics[width=0.3\linewidth]{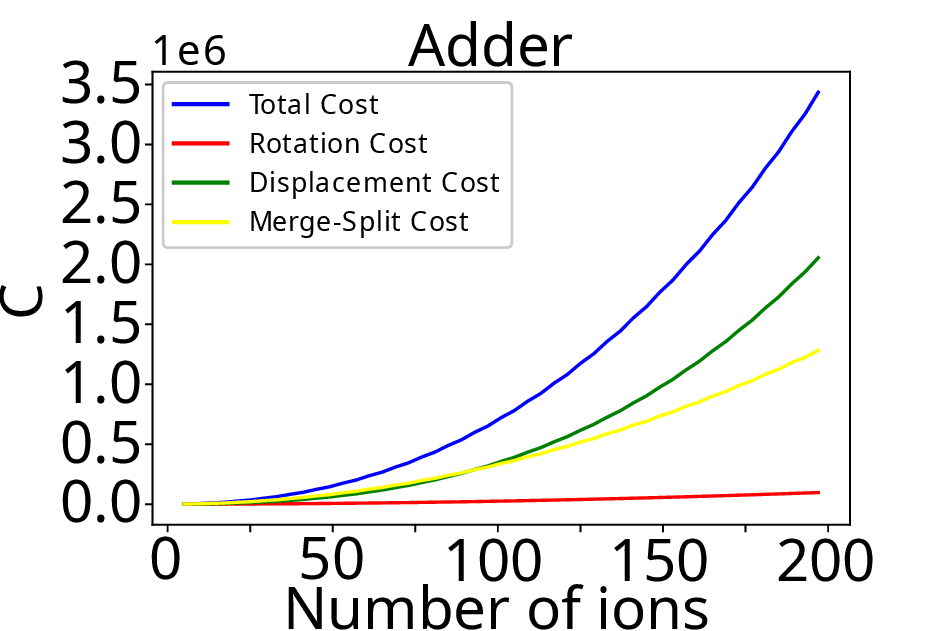}
      \label{subfig:ShiftTotGraph}
    }
    \caption{Results for the CIO algorithm with the cost contribution of each type of operation.
}
    \label{fig:ResultAll}
\end{figure*}

To resolve the issue raised at the end of the previous section, a new shuttling approach is proposed, namely the reorganization of ions while shuttling proceeds. For this, the new algorithm measures the mean crystal distance between interacting ions of the $8$ last executed gates. If that mean distance is above a limit to be called the break distance, then the ions are reorganized to obtain a better placement. When the reorganization is triggered, the algorithm produces a new CIO placement based on the list of gates yet to be executed and then proceeds with this placement to rearrange the ions chain.
The rationale for such a reorganization is that repositioning the ions into a more efficient configuration for further shuttling is less costly than proceeding with a bad positioning.

The break distance is defined  here as the number of crystals present in the circuit (the number of ions divided by two) divided by $12$. The divisor was empirically chosen and can be adapted for testing multiple break distances. As a function of the break distance, the performance of  the reorganization algorithm follows a trend whereby it is inferior to the CIO algorithm without reorganization except for break distance values in the interval $[6,14]$, with the optimum found at approximately~12.

To reorganize the ions, the algorithm uses the CIO to generate a new positioning for the remaining gates to be executed which are considered as a new circuit. 
The ions are then repositioned to their new configuration in the trap by way of shuttling.




This new algorithm gives the circuit fit results appearing in Fig.~\ref{fig:ResultReorg}. It is seen that for the QFT, Carry, Yoyo and Adder circuits the cost is left unchanged by the new shuttling algorithm with reorganization. This shows that the reorganization does not affect already optimal and near-optimal shuttling as the reorganization is not triggered in those cases. The graphs of the results obtained for the Comparator and Shift circuits in Fig.~\ref{fig:ResultReorg}~(e) and~(f) show improvement after around 40 ions. This supports the use of the reorganization procedure for complex circuits, especially for the Shift circuit, the only circuit for which CIO is worse than OAI (see Fig.~\ref{fig:Result}). Thus the reorganization procedure should be used for costly circuits whose circuit fit grows indefinitely for a large number of ions  (as of now around 50 ions).

\section{Displacement cost}
\label{sec:Displ}

In
this
section, the total cost appearing in Eq.~\eqref{eq:CircFit} for the circuit fit will account for the cost of all operations appearing in Table~\ref{tabcost}.
Because the trap is linear and since there is a unique LIZ (case considered thus far), the displacement cost is a direct consequence of the exchange of ions. Hence reducing the number of exchanges reduces the displacement cost. Yet, this cost cannot solely be reduced by the shuttling as the nature of the displacements depends on the architecture itself. Fig.~\ref{fig:ResultAll} shows the results of decomposing the total cost into the different costs associated with each type of operation as per Table~\ref{tabcost}. It is seen that the displacement cost overtakes the exchange cost after a certain number of ions depending on the circuit (except for Carry, but the merge-split cost grows linearly while the displacement cost grows polynomially).
Its reduction thus becomes important for circuits with large numbers of ions. 
Recall, however, that it is not possible to do so solely through the shuttling algorithm, as the displacement problem is inherent to the uni-LIZ topology. This can be proved mathematically as follows.

\begin{figure}[tbp]
\centering
\includegraphics[width=0.45\textwidth]{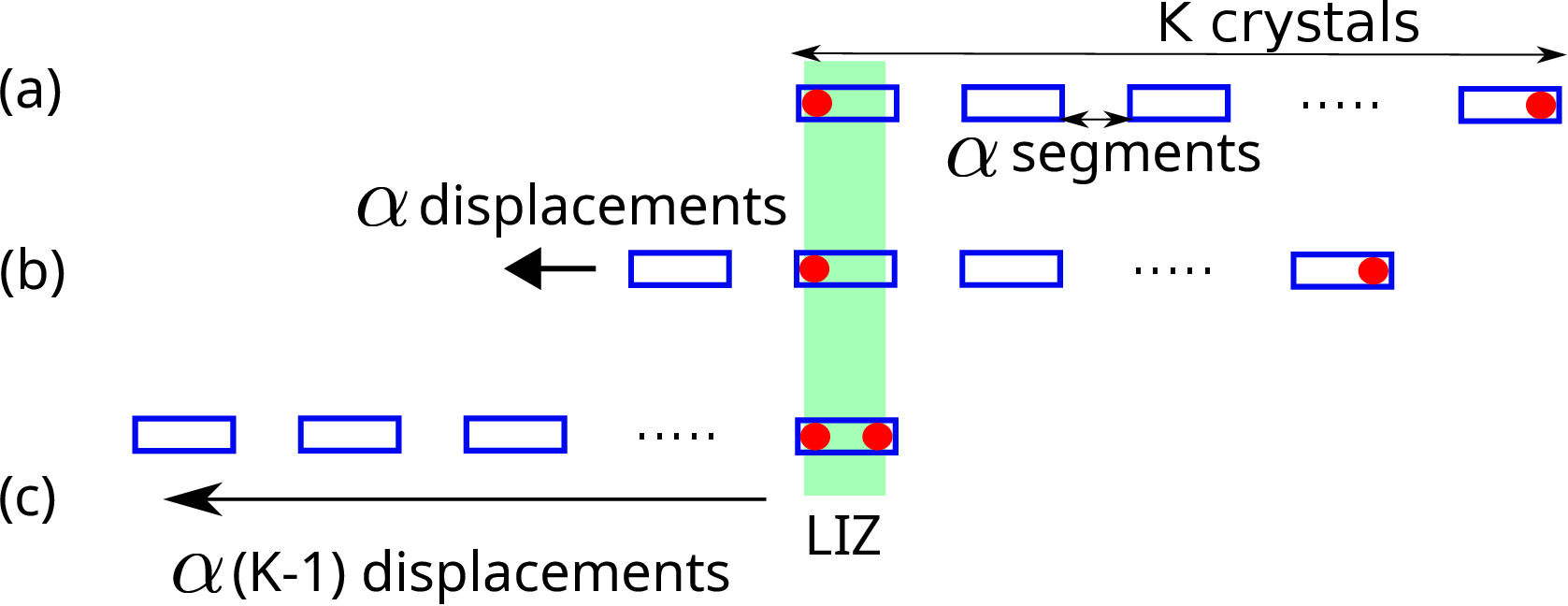}
\caption{Schematic of the worst-case cost for  implementing a two-qubit gate. (a) State at the beginning: The ions, represented as red circles, are at opposite ends of the crystal chain. (b) Result after one ion exchange: One of the crystals has moved by a distance of  $\alpha$ segments.
(c) End result: The $K$ crystals have each moved by a distance $\alpha (K-1)$.}
\label{fig:DisplacementFigure}
\end{figure}

First, the displacement cost of the worst case for implementing a two-qubit gate will be determined. In that case, two ions involved in a gate are at opposing ends of a chain of $K$ crystals. One ion, $I_1$, must be exchanged from one crystal to another until it reaches the second ion $I_2$ in the same crystal. As depicted in Fig.~\ref{fig:DisplacementFigure}~(a), consider that $I_1$'s crystal is already in the LIZ, whereas all other crystals are to the right of $I_1$'s crystal separated from each other by a distance of $\alpha$ segments (minimal number of segments between crystals imposed by the trap physics; $\alpha =1$ for the Mainz computer and this is the value considered here).
For each ion exchange between neighboring crystals to proceed, the two interacting crystals need to move to execute the split and merge operations (Fig.~\ref{FigureCompilerCircuit}). Such displacement comes with a cost. Thus, each ion exchange comes with a displacement cost denoted $P$.
Then, the \textit{leftmost} crystal in the crystal chain that no longer contains $I_1$ must be moved to the left by $\alpha$ segments to make space for the next crystal in order to perform the next exchange. For this, the entire crystal chain is moved to the left as shown in Fig.~\ref{fig:DisplacementFigure}~(b).
In total, for the entire crystal chain, there are $K$ crystal displacements of length $\alpha$. This is repeated $K-1$ times in order to implement the shuttling operation, once for each crystal of the chain minus the first one that was already in position. Thus, the cost for the number of displacements needed to implement a gate is
\begin{equation}
C_g(K) = \alpha K(K-1) + P(K-1).
\label{Eq:costWorse}
\end{equation} 
Here, the cost of a displacement has been set to 1 rather than 2, see Table~\ref{tabcost}; the value of $P$ can be normalized accordingly.
The state shown in Fig.~\ref{fig:DisplacementFigure}~(c) is thus reached.
By the same reasoning, if the second ion $I_2$ is contained in the $K_i^{\mathrm{th}}$ crystal of the chain, the associated cost is given by $C_g(K_i)$.

Eq.~\eqref{Eq:costWorse} gives a quadratic growth of the displacement cost for the worst case and this behavior is conserved for any ion placement.
To further illustrate the inefficiency of the uni-LIZ architecture, it will now be shown that in any case, the displacement cost always grows with the number of ions in such a way that the circuit fit can never be constant.

\begin{figure}[t]
\centering
\includegraphics[width=0.30\textwidth]{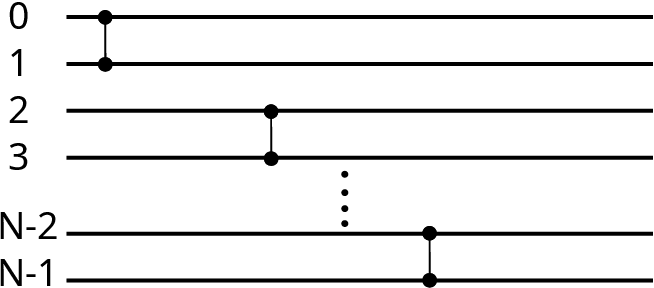}
\caption{Simplest circuit used: each qubit is paired with another one through only one gate.}
\label{fig:Simplest_circuit}
\end{figure}

First, consider the simple circuit illustrated in Fig.~\ref{fig:Simplest_circuit} with $N$ qubits and $\frac{N}{2}$ two-qubit gates. In this circuit, any two qubits only appear in one gate. The optimal placement is obvious to find: the two ions involved in a gate shall be labeled so as to be in the same crystal. There is then no need for ion exchanges between crystals, and shuttling only requires displacements. The optimal placement is to place the crystals by order of gate execution with the first crystal in the LIZ as Fig.~\ref{fig:DisplacementFigure}~(a) shows. The circuit is then implemented by moving the crystal chain from one side to the other. The cost obtained in Eq.~\eqref{Eq:costWorse} is applicable, and using it in the circuit fit definition with $K = \frac{N}{2}$  gives
\[ \mathrm{Circuit~fit} = \frac{\alpha \frac{N}{2}(\frac{N}{2} - 1)}{\frac{N}{2}} = \alpha (\frac{N}{2} - 1).\]
This shows that the cost for this simple circuit grows with the number of ions for a linear uni-LIZ architecture.

This can be generalized to any circuit.
Consider a linear uni-LIZ architecture in which the circuit is executed. Assume also that it is a perfect ion computer with a theoretically optimal shuttling algorithm (in the sense that no better algorithm exists), and it is so good that sequences of one-qubit and two-qubit gates targeting the same list of two qubits can be concatenated in one unique gate implemented without overhead cost. This model computer will be used for the following demonstration.

\begin{figure}[t]
\centering
\includegraphics[width=0.28\textwidth]{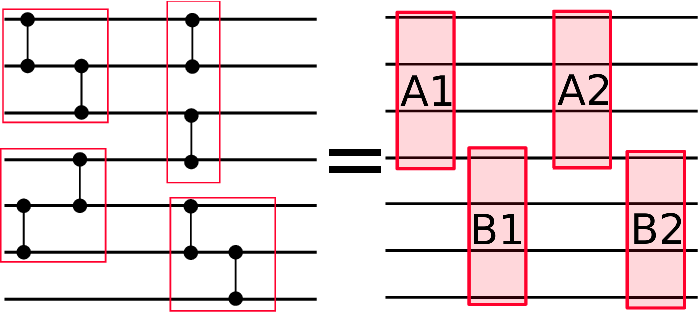}
\caption{A general circuit partitioned into minimal sub-circuits involving two crystals and two-qubit gates.}
\label{fig:GeneralCircuit}
\end{figure}

First, consider a generic circuit $H$ with $N$ qubits and a circuit depth $\geq \frac{N}{2}$.
Only circuits useful for quantum computation will be considered, that is circuits containing two qubit-gates (circuits containing only one-qubit gates are obviously not interesting for quantum computation; evidently the circuit fit that case is 0). 
Such a useful circuit can be partitioned into minimal sub-circuits.
By minimal, here, is understood a sub-circuit with the smallest number of crystals, namely two (four ions), and the smallest number of gates, also two. Such a partition is depicted in Fig.~\ref{fig:GeneralCircuit}.

Because the shuttling algorithm is theoretically optimal, at the execution of a sub-circuit, all ions and crystals are in perfect position without supplementary cost, the first crystal being already in the LIZ and the second crystal being only $\alpha$ segments away from the LIZ. Because of the concatenation in one single gate (see above), the first gate is executed on the first crystal and the second gate on the second crystal with one ion exchange if needed. The structure is the same as the smallest circuit presented above, thus the cost associated with such a sub-circuit is the same as the smallest circuit.

By summing the limit of the circuit fit of every sub-circuit, the minimal theoretical cost for any possible circuit is given by 
\begin{equation}
    \alpha K(K-1) + P(K-1).
    \label{theo_min}
\end{equation}
This demonstrates that the circuit fit for a linear architecture, counting every operation in the total cost, can never be constant versus the number of ions. A linear growth of the circuit fit is the best that can theoretically be achieved.

Furthermore, as the displacement cost per gate cannot be constant, it will become the dominant cost component as the number of ions grows even if the circuit is optimized, for the displacement cost per gate necessarily grows with the number of ions, contrary to other costs such as for merges and splits.

\begin{figure}[tbp]
\centering
    \subfloat{
      \includegraphics[width=0.48\linewidth]{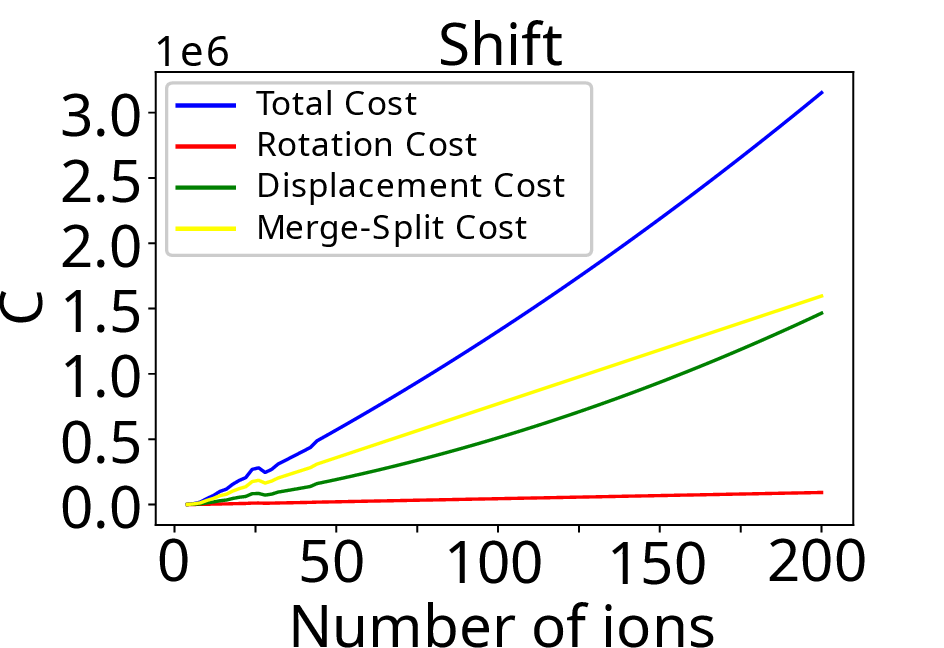}
     \label{subfig:ShiftTotReorgGraph}
    } 
    \subfloat{
      \includegraphics[width=0.48\linewidth]{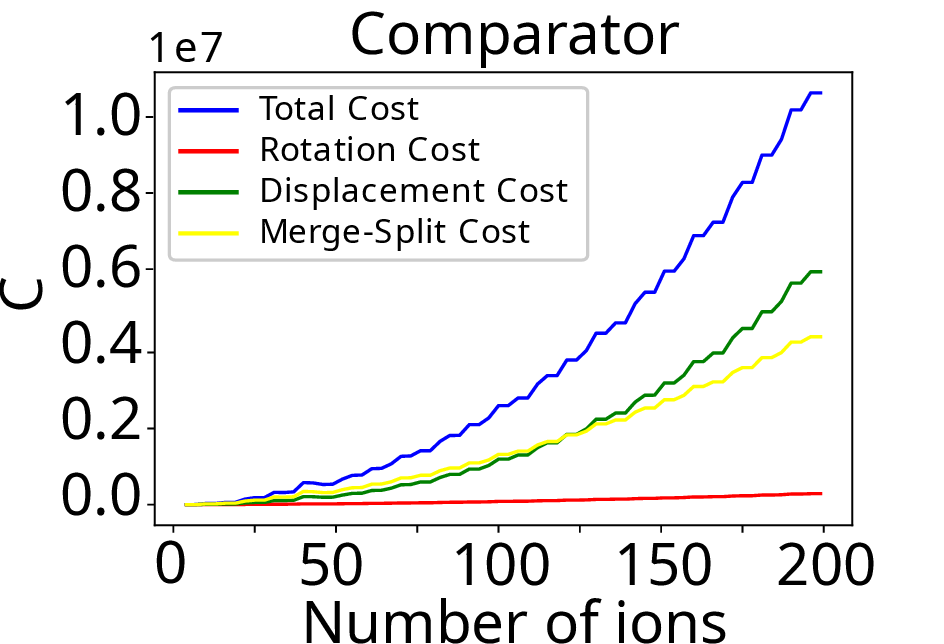}
      \label{subfig:ComparatorTotReorgGraph}
    }
    \caption{Results for the cost with the reorganization algorithm for the Shift and Comparator circuits.
}
    \label{fig:ResultReorgAll}
\end{figure}

It is thus proved that the mean cost per gate will always grow with the number of ions for a uni-LIZ architecture. Hence, the scaling up of the uni-LIZ architecture becomes problematic. For the QFT circuit, whose circuit depth is $\frac{N(N+1)}{2}$, the displacement cost is $\alpha N^3 + P N^2$, that is a cubic growth, while a uni-LIZ architecture is best adapted to the QFT circuit structure~\cite{1article}.
The difficulty in solving the displacement problem lies in the fact that, in a one-dimensional linear architecture, the geometry of the traps does not allow the algorithm to have a direct impact on displacements.
As the expression in Eq.~\eqref{theo_min} shows ($K = \frac{N}{2}$), the  cost is minimally of the order of $N^2$ for a uni-LIZ architecture. Thus, only a reduction of the number of ion exchange operations opens the possibility to reduce the displacement cost.
As Figs.~\ref{fig:ResultAll} and~\ref{fig:ResultReorgAll} show, as the number of ions grows the displacement cost grows beyond the split/merge cost in every case (except for Carry), even if reorganization reduces the different cost components (by a reduction factor of 10 for the Shift circuit). Scaling of the uni-LIZ architecture is not a physical possibility.

Because the displacement problem is inherent to the linear uni-LIZ architecture, there is thus the need to explore a multi-LIZ architecture.

\section{Multi-LIZ architecture}
\label{sec:ML}

\begin{figure}[htbp]
\centering
\includegraphics[width=0.48\textwidth]{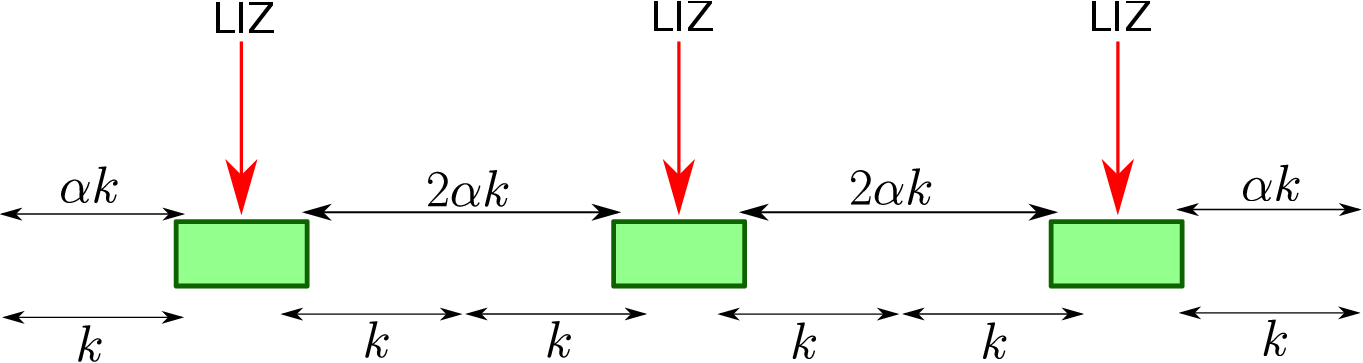}
\caption{
        Multi-LIZ architecture with $L = 3$ LIZ. The LIZ are separated from each other by a distance of $2 \alpha k$ segments, each LIZ section storing up to $k$ crystals on each side of the LIZ.}
\label{fig:MULTIArchi}
\end{figure}

\begin{figure*}[htbp]
\centering
    \subfloat{
      \includegraphics[width=0.3\linewidth]{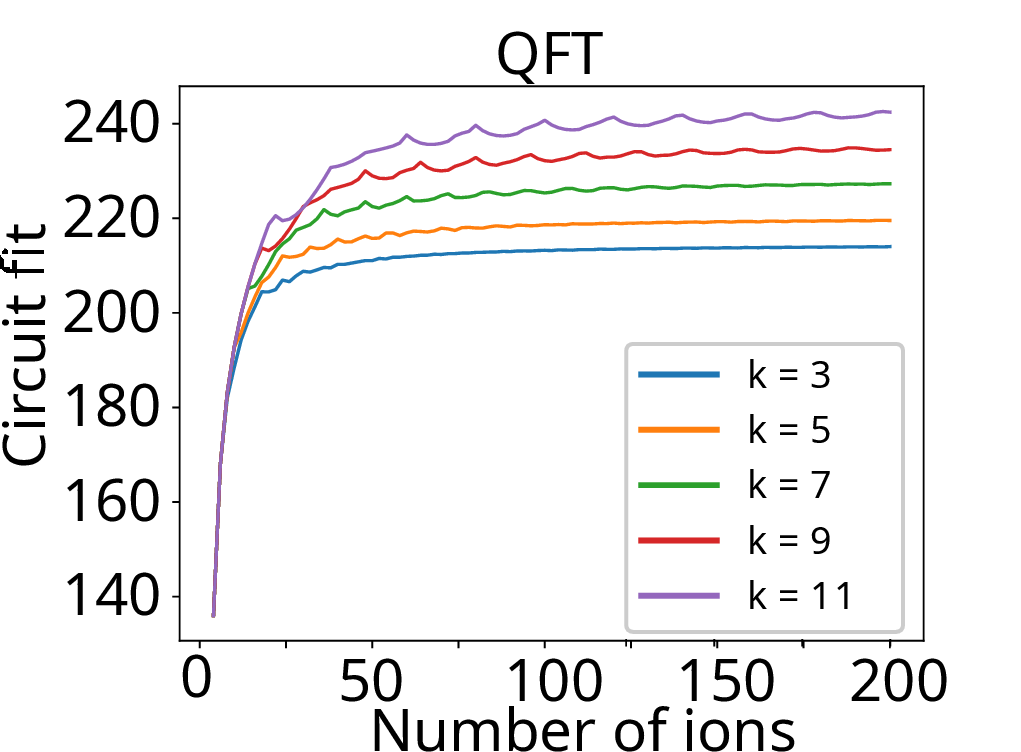}
      \label{subfig:QFTMLGraph}} 
    \subfloat{
      \includegraphics[width=0.3\linewidth]{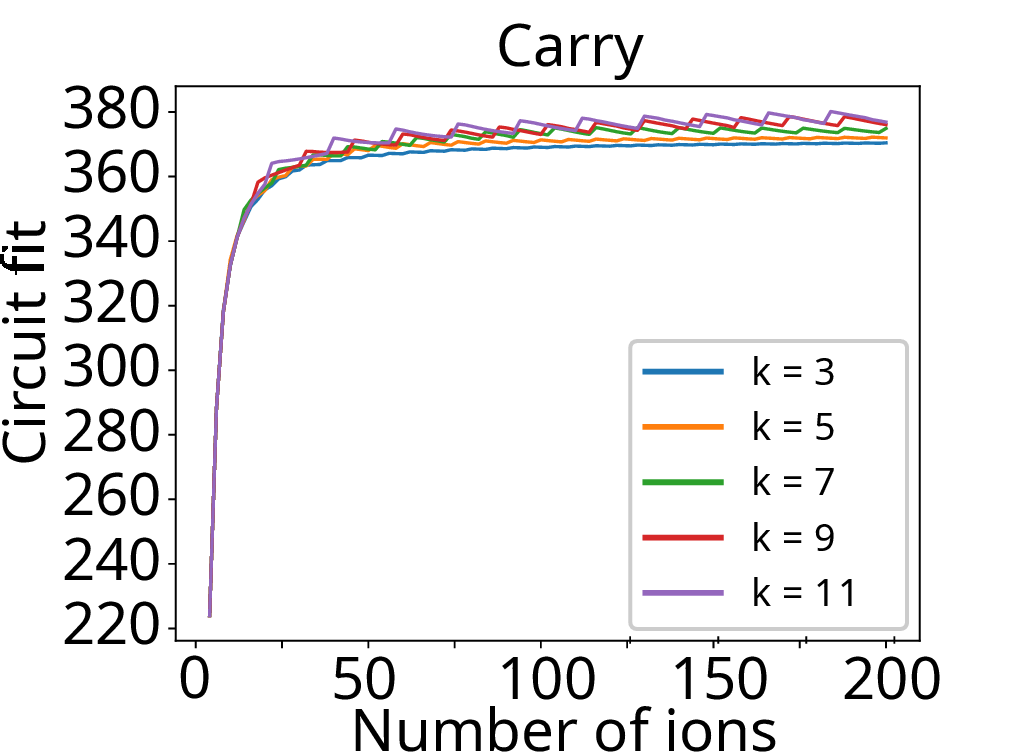}
      \label{subfig:CarryMLGraph}} 
    \subfloat{
      \includegraphics[width=0.3\linewidth]{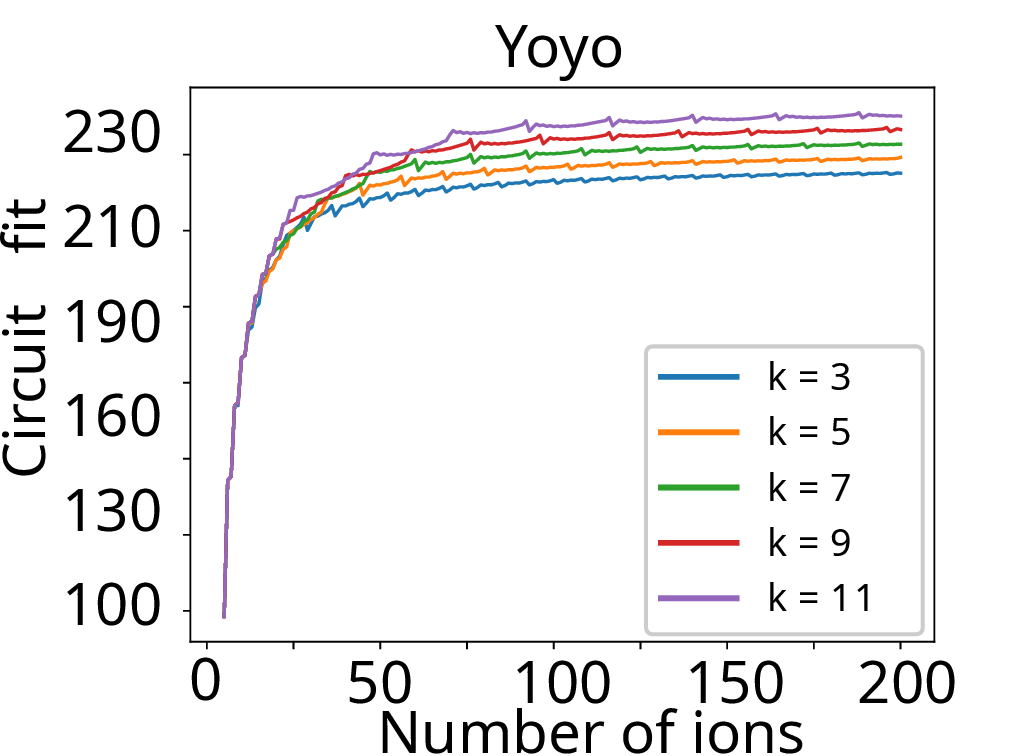}
      \label{subfig:YoyoMLGraph}} 
    \\
    \subfloat{
      \includegraphics[width=0.3\linewidth]{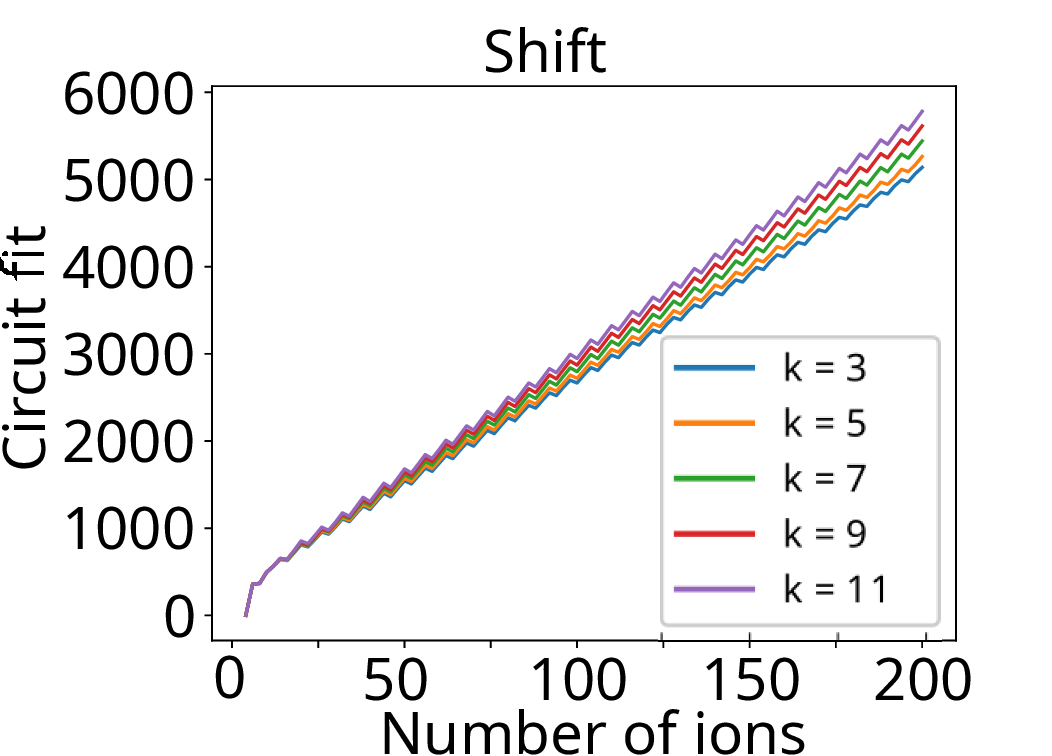}
      \label{subfig:ShiftMLGraph}}
    \subfloat{
      \includegraphics[width=0.3\linewidth]{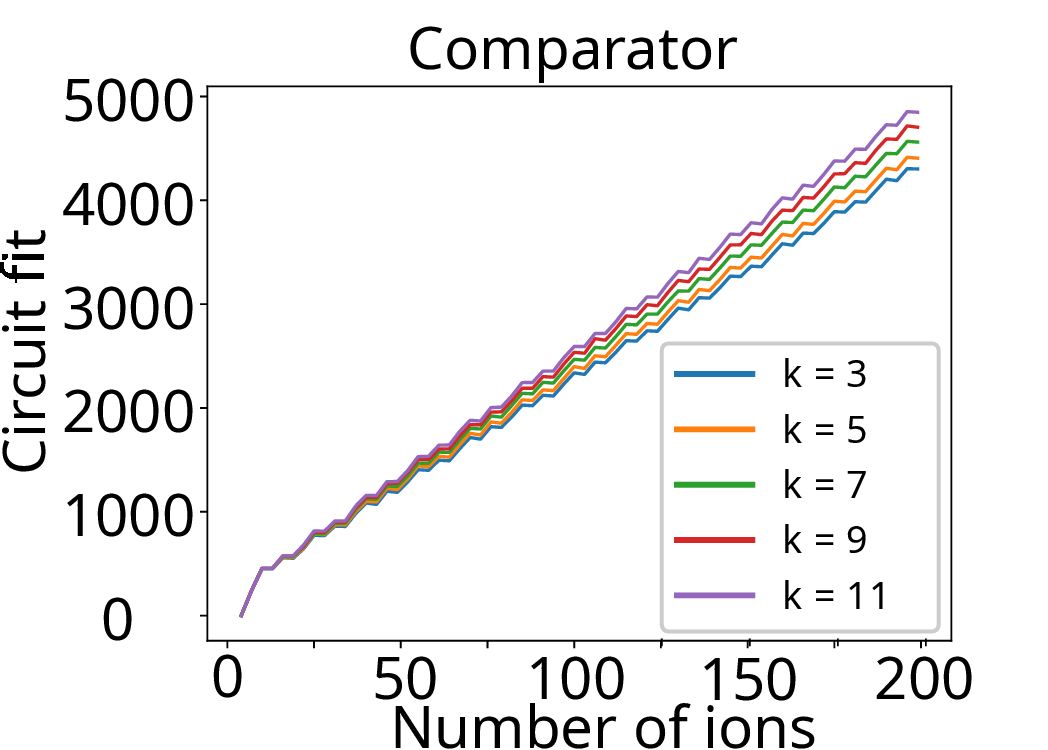}
      \label{subfig:ComparatorMLGraph}}
    \subfloat{
      \includegraphics[width=0.3\linewidth]{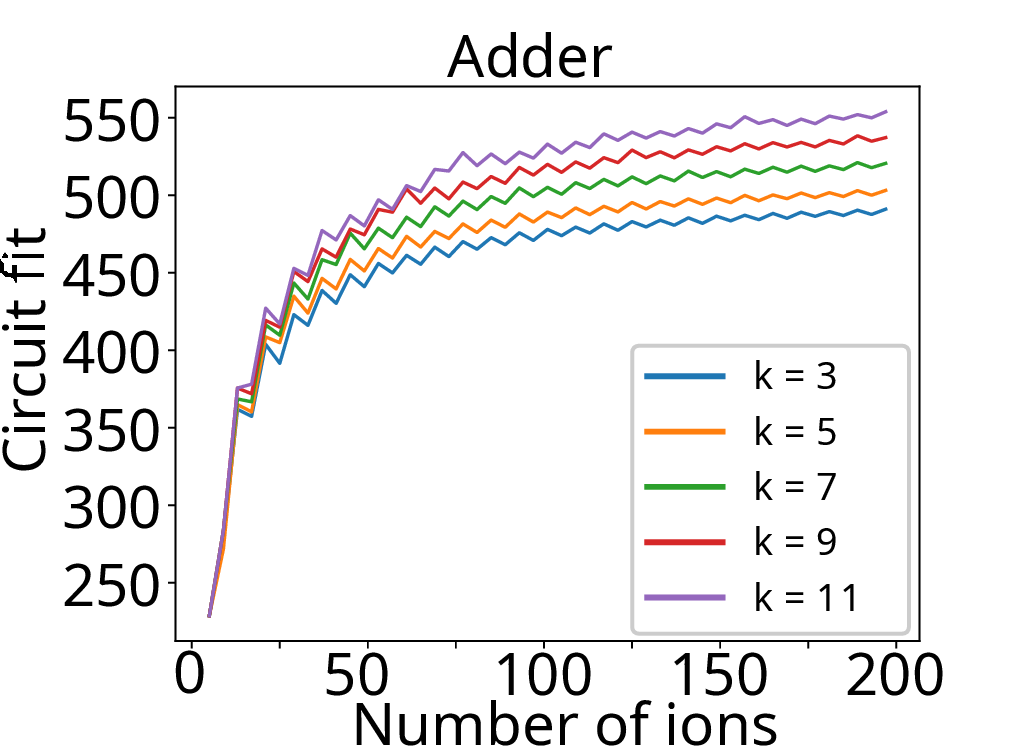}
      \label{subfig:AdderMLGraph}}
    \caption{Results for multi-LIZ linear architectures, with $k$ varying from 3 to 11.}    
    \label{fig:ResultML}
\end{figure*}

This section studies a possible solution that may already be realized with the actual state of technology, that is a linear architecture with multiple LIZ, whereby the linear trap is partitioned into sections each having a LIZ. Each section can contain up to $k$ crystals on each side of its LIZ, and it is assumed that each section has enough segments, so that every crystal can be shuttled without interacting with other LIZ's crystals. The architecture is depicted in Fig.~\ref{fig:MULTIArchi}. Let $L$ be the number of LIZ and $k = \frac{K}{L}$, with $K$ the total number of crystals; $k$ is the average number of crystal per segment.

A formula  will now be derived for the cost of the worst case for a multi-LIZ architecture, whereby the targeted ions are in separate crystals at the opposite ends of the segmented trap.

The cost of shuttling in one section is first calculated. Using Eq.~\eqref{Eq:costWorse}, the shuttling cost of bringing one ion from one end to another of one LIZ section is given by $\alpha k(k-1) + P(k-1).$
Now, the ion must go from one LIZ section to another LIZ section. Because of this, a cost $2 \alpha (K-1) + P(L-1)$ is added. The worst case occurs when one ion must permute across the crystal chain from one end of the trap to the other; the cost is then
\begin{equation}
\begin{split}
    C =\,& 2 \alpha (K-1) + P(L-1) \\ &+ \sum^L_{l=1}{ \alpha k(k-1) + P(k-1)}.
\end{split}
    \label{totalcost_EQ}
\end{equation}
Here, $2 \alpha (K-1) + P(L-1)$ is a constant that depends solely on the architecture topology.
Since the quantity in the summation does not depend on $l$,
and using $k = \frac{K}{L}$, the previous equation becomes
\begin{equation}
C = 2 \alpha (K-1) + \alpha K(\frac{K}{L}-1) + P(K-1).
\label{eq_last_cost}
\end{equation}
From Eqs.~\eqref{Eq:costWorse} and~\eqref{eq_last_cost},  the following inequality holds if the multi-LIZ architecture is to be better than the uni-LIZ:
\begin{equation}
\begin{split}
2 \alpha (K-1) + \alpha K(\frac{K}{L}-1)  + P(K-1) \\\le \alpha K(K-1) + P(K-1), 
\end{split}
\end{equation}

and thus
\[2 (K-1) +K(\frac{K}{L}-1) - K(K-1) \le 0,\]
from which it is deduced that
$L \ge 2$ if the multi-LIZ is to be better than the uni-LIZ $\forall K \in \mathbb{N}$.

As the number of LIZ grows, Eq.~\eqref{eq_last_cost} shows that the cost still depends quadratically on $K$, but is greatly attenuated if the number of LIZ is large. Thus, more LIZ reduces the displacement cost.

To validate the results obtained above numerically, simulations have been carried out with different values of $k$ in an architecture with a thousand segments. For each $k$, the architecture is partitioned into enough LIZ that the distance between each LIZ is $2 \alpha k$. The shuttling algorithm uses the CIO heuristic discussed previously with some modification of implementation (but not of logic) to adapt it to the multi-LIZ architecture. The initial positioning of crystals is no longer by segments but by LIZ sections. The crystals are distributed LIZ section by LIZ section until each section is full (with $k$ crystals per section).
The better results come from the cost reduction brought about by the reduced number of crystals in a section; results are thus presented according to the value of $k$, see Fig.~\ref{fig:ResultML} which is to be compared with Figs.~\ref{fig:Result} and~\ref{fig:ResultReorg}. The following observations can be made:

\begin{itemize}
    \item The linear multi-LIZ architecture developed here allows mitigating the explosion of the displacement cost for the QFT, Carry, 
    Yoyo, and Adder (Fig.~\ref{fig:ResultML}~(a), (b), (c) and (f)) as the circuit fit asymptotically follows a linear growth with a smaller coefficient for large numbers of ions. As $k$ diminishes, the circuit fit tends to a vertical asymptote, but never converges to a constant.
    \item Less optimized circuits such as Shift and Comparator (Fig.~\ref{fig:ResultML}~(d) and~(e)) see a large cost reduction of 60\% for the Shift circuit and 40\% for the Comparator circuit. There is thus an improvement for all circuits even for large $k$. 
    \item The oscillations appearing in the results are caused by the saturation of LIZ sections as the number of ions grows. As $k$ diminishes, so do the oscillations' amplitude and period.
    \item The cost reduction decreases with each $k$ (recall that the smaller $k$ is, the larger the number of LIZ $L$). Thus, at a certain point, adding more LIZ has a decreasing return.
\end{itemize}
 
The previous observations demonstrate the theoretical and practical interest of a multi-LIZ architecture. Even if the circuit fit asymptotically becomes a straight line as Fig.~\ref{fig:ResultML} shows, a multi-LIZ architecture
reduces the cost of shuttling. However, the cost of a larger number of LIZ may prove to be prohibitive as the number of qubits grows.
Investigating shuttling operations across X-junctions will open two-dimensional architectures~\cite{conta2026toolchainshuttlingtrappedionqubits}.

\section*{Acknowledgements}

J.D. acknowledges financial support from the QSciTech training program funded by the Collaborative Research and Training Experience (CREATE) program of the Natural Sciences and Engineering Research Council of Canada (NSERC). Y.B.L. acknowledges support from the Canada First Research Excellence Fund (CFREF) through the Institut quantique at Universit\'{e} de Sherbrooke. U.P. and F.S.K. acknowledge funding by the German Bundesministerium für Forschung, Technologie und Raumfahrt (BMFTR) within the projects SYNQ and ATIQ, and by the Deutsche Forschungsgemeinschaft (DFG) within the project comfortQC in the SPP2514. The research is based upon work supported by the Office of the Director of National Intelligence (ODNI), Intelligence Advanced Research Projects Activity (IARPA), via the U.S. Army Research Office grant W911NF-16-1-0070. The views and conclusions contained herein are those of the authors and should not be interpreted as necessarily representing the official policies or endorsements, either expressed or implied, of the ODNI, IARPA, or the U.S. Government. The U.S. Government is authorized to reproduce and distribute reprints for Governmental purposes notwithstanding any copyright annotation thereon. Any opinions, findings, and conclusions or recommendations expressed in this material are those of the author(s) and do not necessarily reflect the view of the U.S. Army Research Office.

\begin{IEEEbiography}
[{\includegraphics[width=1in,height=1.25in,clip,keepaspectratio]{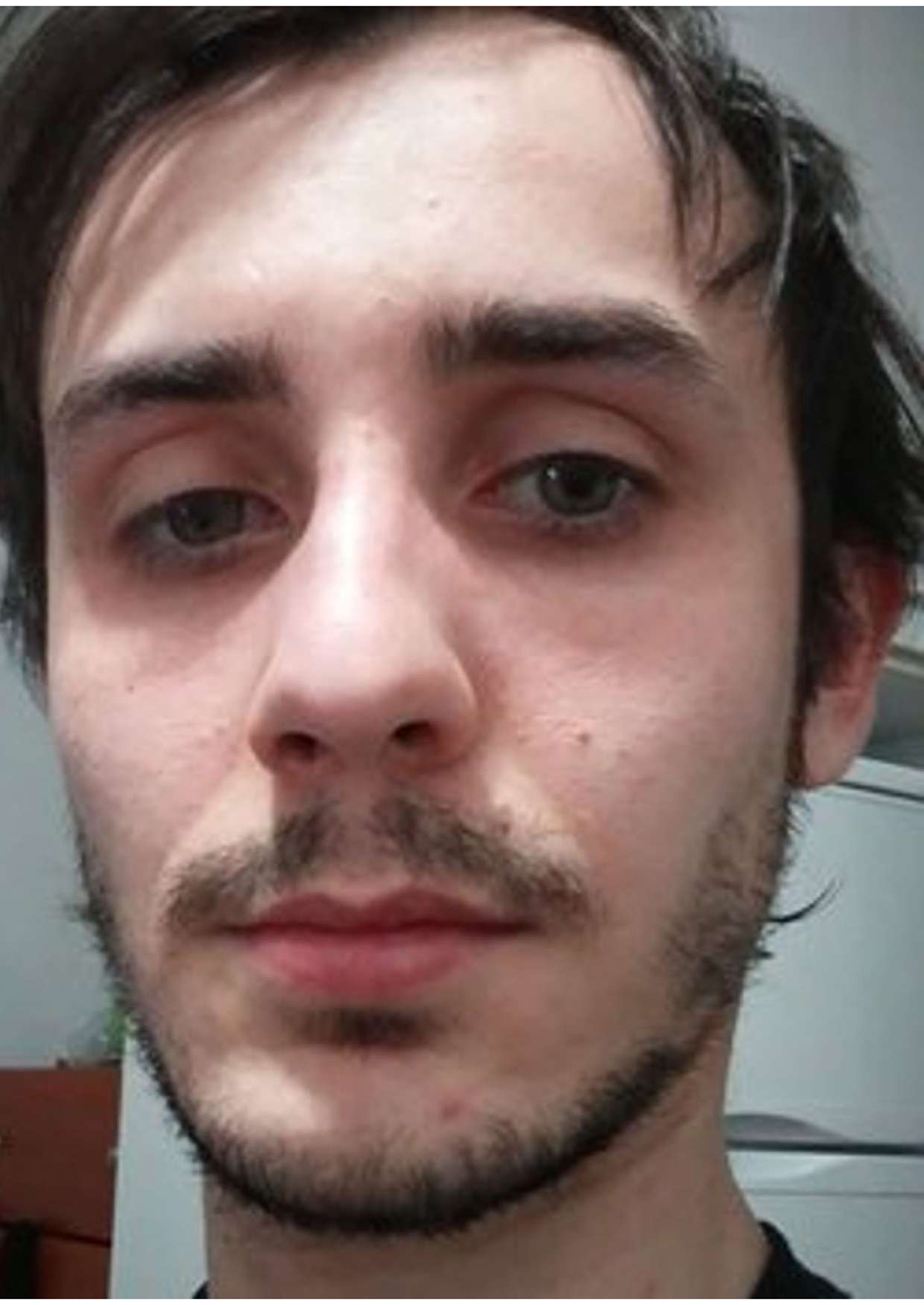}}]{Jonathan Durandau}, M.Sc.A is a Ph.D candidate in electrical engineering at Université de Sherbrooke, and a student member Institut quantique (IQ), who specializes in quantum engineering. His research interests include quantum circuit analysis, trapped ion shuttling, and compilers for quantum computers.
\end{IEEEbiography}

\begin{IEEEbiography}
[{\includegraphics[width=1in,height=1.25in,clip,keepaspectratio]{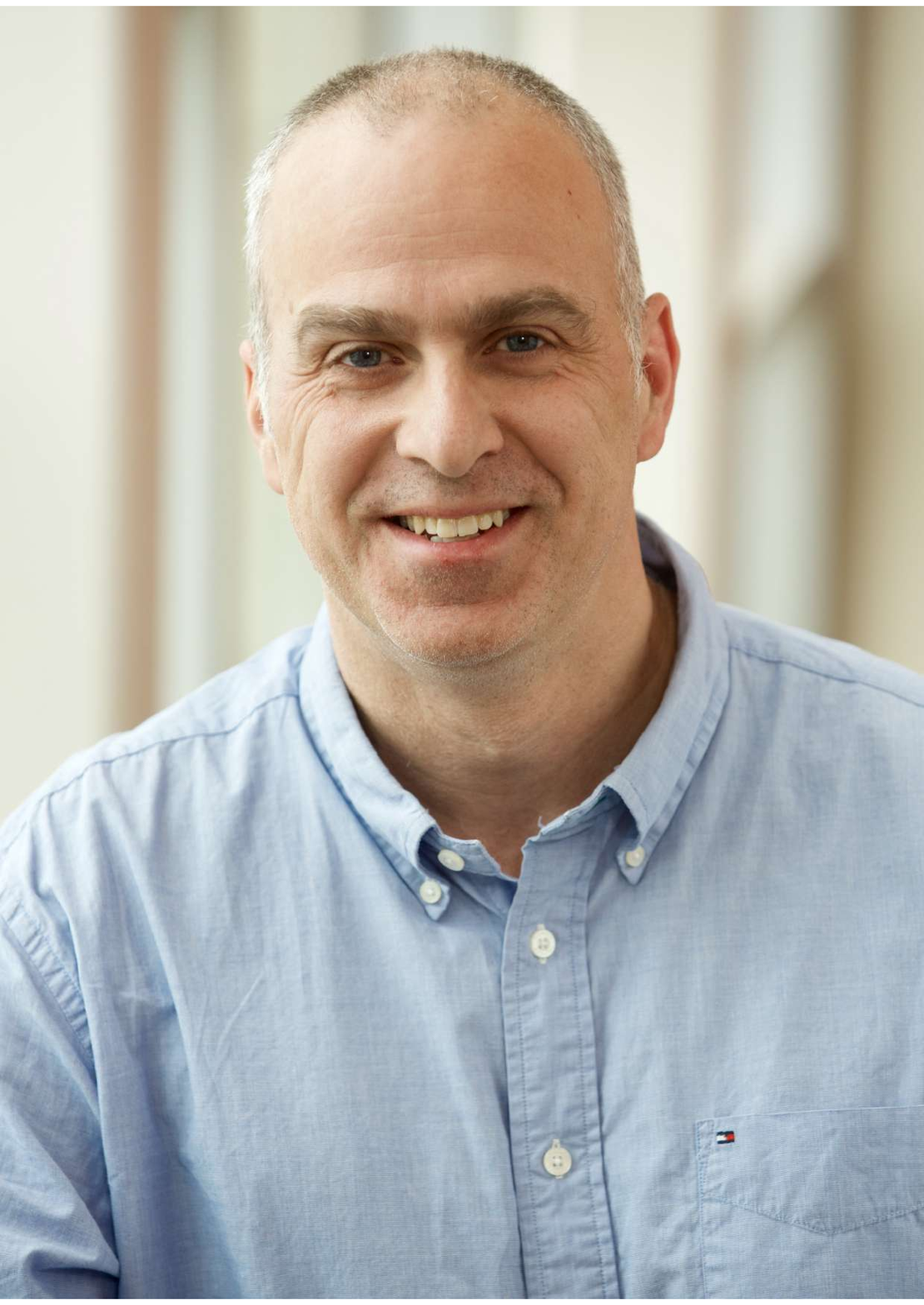}}]{Charles-Antoine Brunet}, P. Eng, Ph.D. is a computer engineer and professor of electrical and computer engineering at the Université de Sherbrooke (UdeS) since 2000. At UdeS, he has also served as the academic director of the computer engineering bachelor's degree program (2009-2019 and 2025 to now) and vice-dean in informational resources (2021-2025). His research interests are in computer programming and languages, applied artificial intelligence (aerospace, music, positron emission tomography scanners), and quantum computing.
\end{IEEEbiography}

\begin{IEEEbiography}
[{\includegraphics[width=1in,height=1.25in,clip,keepaspectratio]{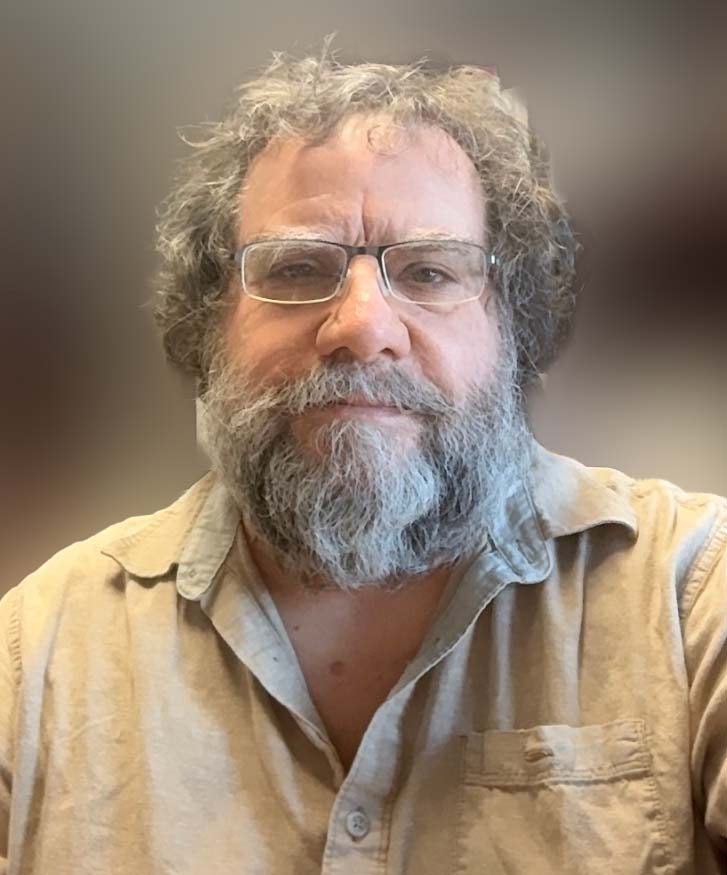}}]{Frédéric Mailhot}, P.Eng., PhD. is an engineering physicist and professor of electrical and computer engineering at Université de Sherbrooke since 1995.  Prior to coming to Sherbrooke he was a Senior computer engineer at Synopsys.  At UdeS, he has served as the academic director of the electrical engineering bachelor’s degree program (2008-2012), Head of the electrical and computer engineering department (2012-2017), academic director of the electrical engineering professional master’s degree program (2017-present). His research interests are in  computer security, computer architecture and synthesis (both for classical and quantum systems) and pedagogy.
\end{IEEEbiography}

 \begin{IEEEbiography}[{\includegraphics[width=1in,height=1.25in,clip,keepaspectratio]{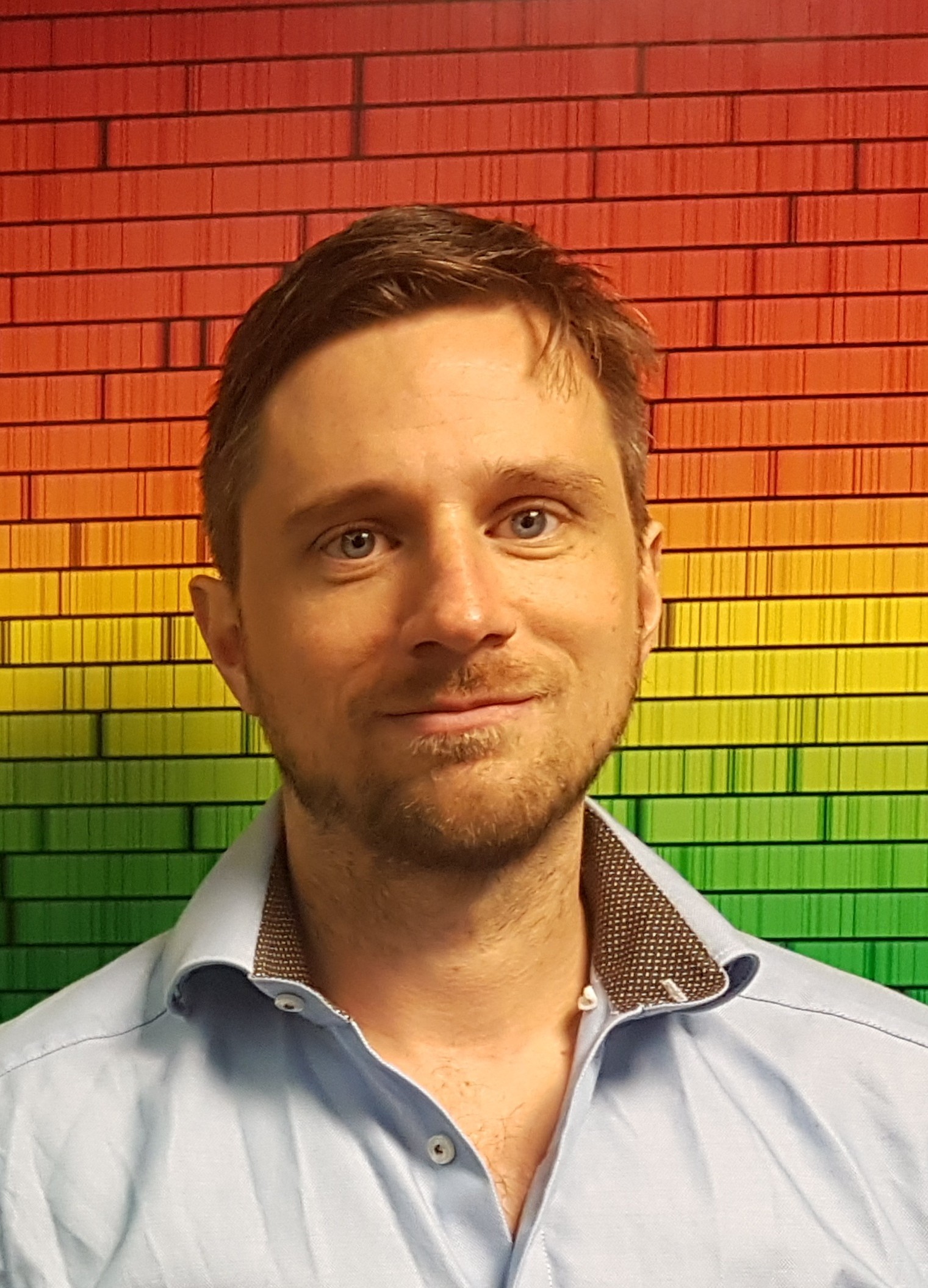}}]{Ulrich Poschinger}, Dr., is a physicist. He acts as assistant professor in experimental physics at the Johannes Gutenberg University of Mainz, Germany since 2010. His research interests are in trapping, cooling and manipulating quantum systems, especially working with trapped ions for quantum computing and quantum optics. Also, he has previous experience with neutral atoms.
\end{IEEEbiography}

\begin{IEEEbiography}[{\includegraphics[width=1in,height=1.25in,clip,keepaspectratio]{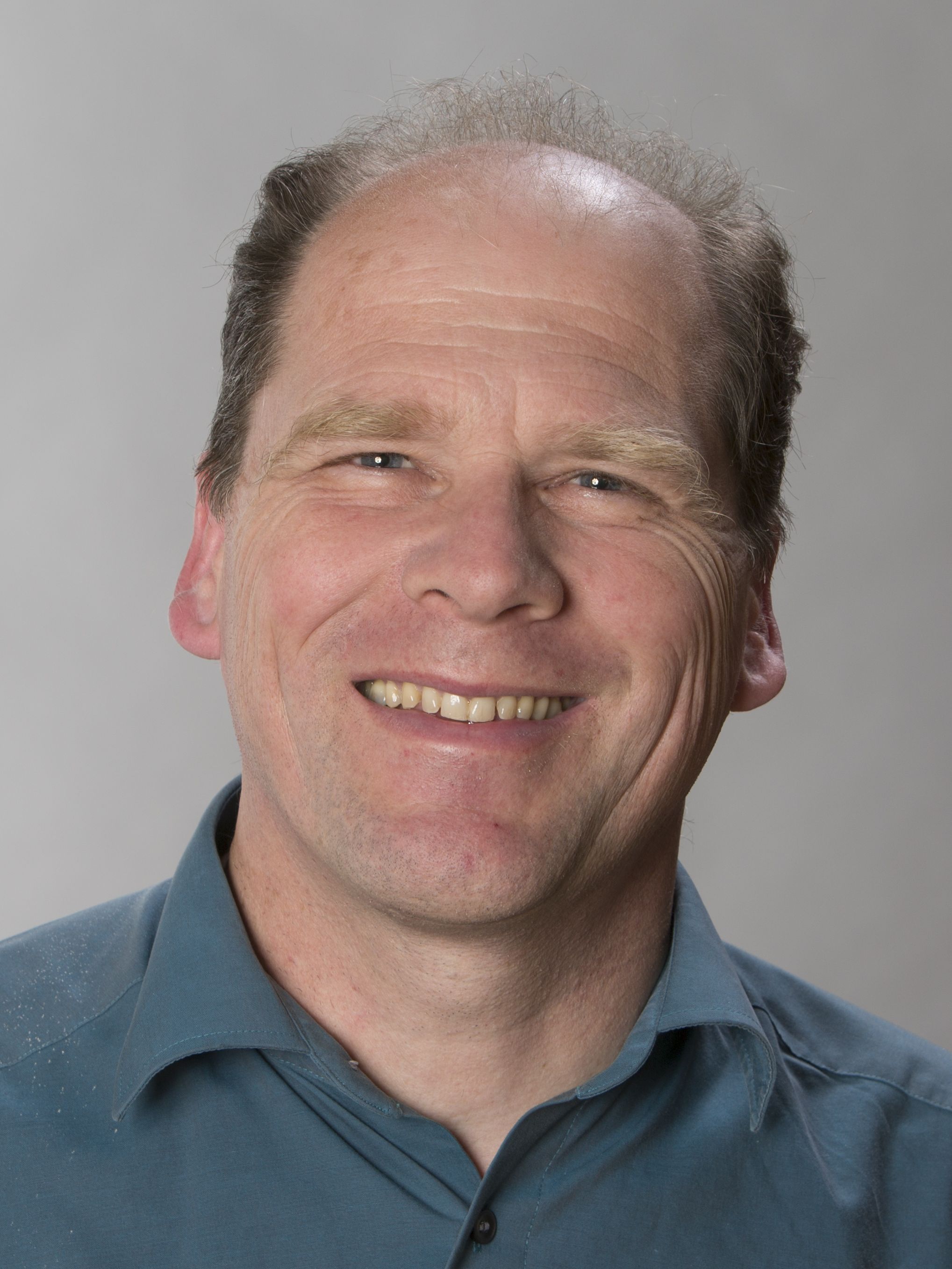}}]{Ferdinand Schmidt-Kaler}, Prof. Dr., is a physicist. He is full professor of experimental physics at the Johannes Gutenberg University of Mainz, Germany since 2010. He is also head of the QUANTUM group at the faculty of physics. Prior to this, he had been full professor at the University of Ulm, Germany. His research interests are in trapping, cooling and manipulating quantum systems, especially working with trapped ions for quantum computing and quantum optics. He also has experience with neutral atoms and ions in Rydberg states, color centers and electrons. He received the Rudolf-Kaiser-Award, and the Helmholtz-Award.
\end{IEEEbiography}

\begin{IEEEbiography}[{\includegraphics[width=1in,height=1.25in,clip,keepaspectratio]{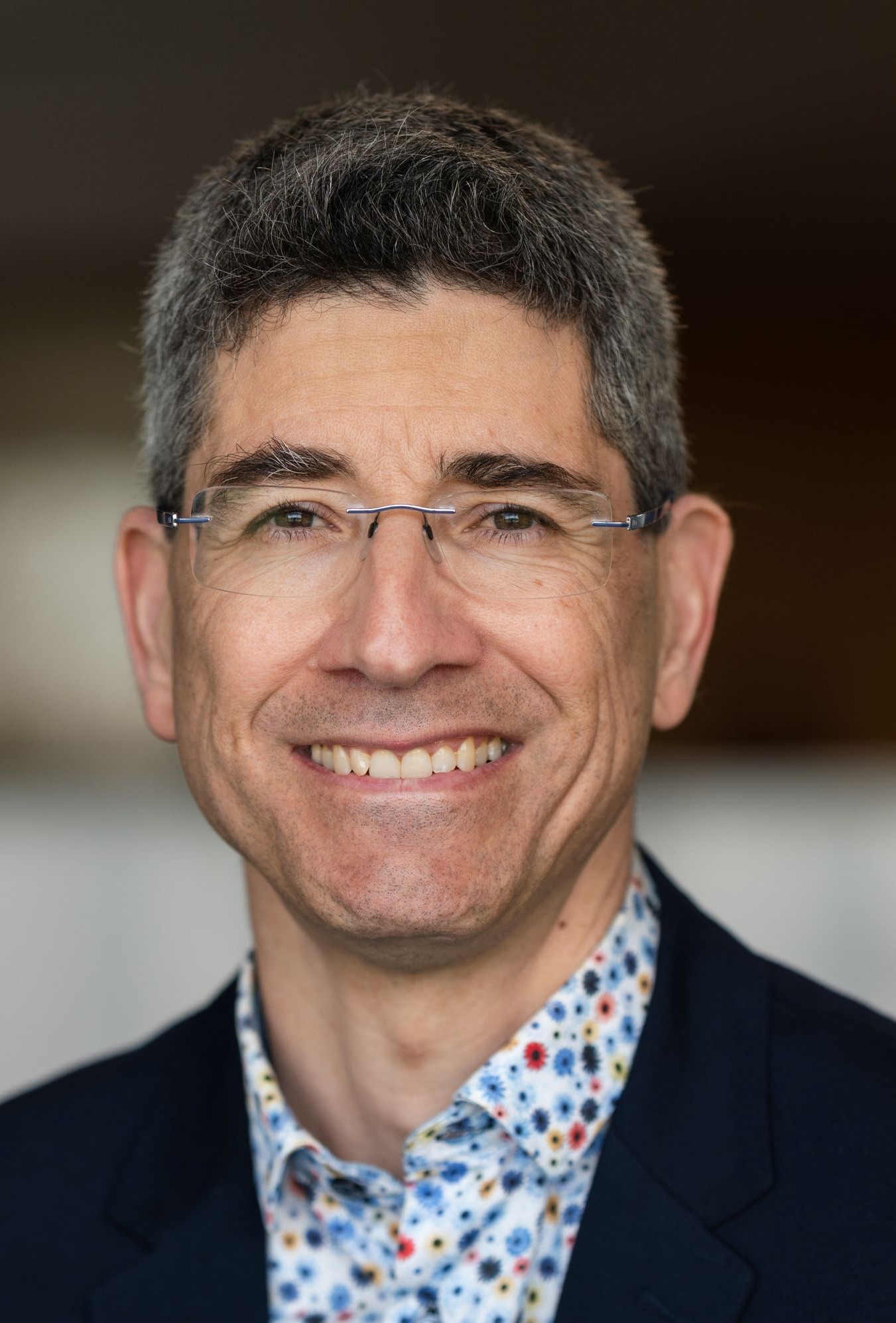}}]{Yves B\'erub\'e-Lauzi\`ere}, P. Eng, Ph.D. is an engineer and physicist. He is professor of electrical and computer engineering at Université de Sherbrooke (UdeS) since 2003 and researcher at Institut quantique (IQ). Prior to coming to Sherbrooke, he was Head of the biophotonics group at Institut national d'optique (INO - Québec City). At UdeS, he has served as the academic director of the electrical engineering bachelor's degree program (2012-2015). From 2018 to 2025, he was the director of the NSERC-CREATE-funded QSciTech training program that he set up. QSciTech provided added-value training in engineering design, project management, teamwork, intellectual property, and entrepreneurship to research students in quantum technologies to better prepare them for a professional career. His research interests are in optimal and feedback control of quantum systems, quantum computing, inverse problems, image reconstruction and instrumentation for diffuse optical tomography and X-ray computed tomography.
\end{IEEEbiography}

\end{document}